\documentclass[10pt,journal]{IEEEtran}

\usepackage[ruled,vlined]{algorithm2e}

\usepackage{bm} % Math packages

\usepackage{epsfig}
\usepackage{amssymb} 
\usepackage[tbtags]{amsmath} 
\usepackage{latexsym}
\usepackage{euscript}
\usepackage{graphics,eepic,epic,psfrag}
\usepackage{enumitem}
\usepackage{xhfill}
\usepackage{relsize}
\usepackage{exscale}

\usepackage[font = footnotesize]{caption}
\usepackage{subcaption}
\usepackage{graphicx}
\usepackage{cite}  
\usepackage{xpatch}
\usepackage{algorithmicx}

\newcommand\norm[1]{\|#1\|}
\newcommand{\RN}[1]{%
  \textup{\uppercase\expandafter{\romannumeral#1}}%
}

\begin{document}
\title{Learning Based User Scheduling in Reconfigurable Intelligent Surface Assisted Multiuser Downlink 
}
\author{Zhongze~Zhang,~\IEEEmembership{Student Member,~IEEE,}
        Tao~Jiang,~\IEEEmembership{Student Member,~IEEE,}
        and~Wei~Yu,~\IEEEmembership{Fellow,~IEEE}% <-this % stops a space
\thanks{The authors are with The Edward S. Rogers Sr. Department
of Electrical and Computer Engineering, University of Toronto, Toronto,
ON M5S3G4, Canada. E-mails: ufo.zhang@mail.utoronto.ca, 
taoca.jiang@mail.utoronto.ca, weiyu@ece.utoronto.ca.
The materials in this paper have been presented in part at the IEEE International Conference on Acoustics, Speech and Signal Processing 
(ICASSP), 2022\cite{zhongzeicassp}.
This work is supported by Huawei Technologies Canada Ltd. Co. 
}% <-this % stops a space
%\thanks{J. Doe and J. Doe are with Anonymous University.Email:\{ufo.zhang, taoca.jiang\}@mail.utoronto.ca, weiyu@ece.utoronto.ca}% <-this % stops a space
}

\maketitle

\begin{abstract}
Reconfigurable intelligent surface (RIS) is capable of intelligently
manipulating the phases of the incident electromagnetic wave to improve the wireless
propagation environment between the base-station (BS) and the users.
This paper addresses the joint user scheduling, RIS configuration, and BS
beamforming problem in an RIS-assisted downlink network with limited pilot
overhead. We show that graph neural networks (GNN) with permutation invariant
and equivariant properties can be used to appropriately schedule users and to
design RIS configurations to achieve high overall throughput while accounting
for fairness among the users. As compared to the conventional methodology of
first estimating the channels then optimizing the user schedule, RIS
configuration and the beamformers, this paper shows that an optimized user
schedule can be obtained directly from a very short set of pilots
using a GNN, then the RIS configuration can be optimized using a second GNN, 
and finally the BS beamformers can be designed based on the overall effective 
channel. Numerical results show that the proposed approach can utilize 
the received pilots more efficiently than the conventional channel estimation based
approach, and can generalize to systems with an arbitrary number of users.
%It can also be applied to RIS with discrete phase shifts.
\end{abstract}

\begin{IEEEkeywords}
Deep learning, graph neural network, reconfigurable intelligent surface, proportional fairness, scheduling.\end{IEEEkeywords}

% \fi
\IEEEpeerreviewmaketitle

\section{Introduction}
\IEEEPARstart{R}{}econfigurable intelligent surface (RIS) is envisioned as a key enabling
technology for a smarter radio environment \cite{basar2019wireless,bible,4fuliu,liasko,coding_metasurface}, due to its capability
to manipulate the phases of wireless signals to enhance the transmission
environment and to improve the network utility (e.g., sum rate
\cite{9lyu2020hybrid} or minimum rate \cite{2qua,irs_perfcsi2,minrate_3}).  This paper
addresses a key problem of user scheduling for the RIS-assisted wireless
cellular network.  In a downlink transmission environment with a base-station
(BS) equipped with $M$ antennas, which can serve at most $M$
users simultaneously over the same resource block, if the total number of users
in the network $K$ is greater than $M$, how should the BS optimally schedule
a subset of users at each timeslot in conjunction with the optimal RIS
configuration and the BS beamforming to achieve high network
throughput while ensuring fairness across the users?

The conventional approach to user scheduling typically follows a two-step
approach of aiming to accurately recover the channel state information (CSI) in
the first step, then based on the estimated CSI, optimizing downlink scheduling in
the second step.  However, the two-step approach is not necessarily optimal as
it suffers from three major shortcomings. First, accurate recovery of a large 
number of unknown channel coefficients is costly in terms of pilot training
overhead. The number of channel coefficients that need to be estimated scales
with the size of RIS, which typically consists of hundreds of passive elements.
Pilot training overhead occupies resources within the channel coherence period
that could otherwise be used for data transmission. Second, the channel 
estimation process typically aims to recover the channel according to some arbitrary metric, e.g.,
mean squared error, which may not necessarily align with the ultimate communication objective. 
Finally, even if the CSI is known perfectly, the user achievable rates are
nonconvex functions of the RIS configuration and BS beamformers; further, the
scheduling problem is discrete. So, the overall problem is a mixed discrete and 
continuous non-convex optimization problem, for which finding an optimal
solution is computationally intensive and difficult to achieve in a real-time
application.

To address these challenges, recent works have advocated learning based approaches to directly map the received pilot sequences to the RIS configuration and the beamformers that maximize the network objective \cite{taojournal,emil2020deep}, in effect bypassing explicit channel estimation. However, the discrete optimization problem of user scheduling, an important but challenging part of the overall network optimization, is not considered in these works. %Therefore, we propose to address user scheduling problem in dense RIS-assisted network by adopting a machine learning approach, to bypass explicit channel estimation and to directly design the schedule, RIS reflection coefficients and beamformer mainly based on received pilot sequences. 

%propose to adopt machine learning approach to bypass explicit channel estimation and to directly design the schedule, RIS reflection coefficients and beamformer to optimize the network objective. 

This paper focuses on the scheduling aspect of an RIS-assisted multiuser
downlink network. Toward this end, we first make an observation that the
overall optimization problem has the following permutation invariant and
equivariant properties: if the ordering of users is permuted, the same
subset of users should be scheduled and the same reflection coefficients should
be used (i.e., permutation invariance), while the beamforming vectors should be
permuted in the same way (i.e., permutation equivariance). This motivates us to make
use of permutation invariant and equivariant graph neural networks (GNN) to
learn the mapping from the received pilots and the user weights to the
optimized RIS configuration and beamforming vectors. This is a generalization
of the neural network architecture proposed in the previous work \cite{taojournal}, which deals only with beamforming and reflective coefficients design, to
a system-level optimization that also accounts for the user priorities in the
scheduling process. Scheduling in a multiuser network is in general 
challenging to learn, because the scheduling choice can be highly sensitive to 
the user priority weights. In this paper, we advocate the following three-stage
system-level optimization approach. Specifically, we show that in the first 
stage, a GNN applied to all potential users but with very short pilots can already
produce an optimized schedule while accounting for the user priorities. This is
achieved by adopting an \emph{implicit scheduling} scheme and by inferring the
schedule from the set of output beamforming vectors. In the second stage, a
second GNN applied only to the scheduled users but with longer pilots is used
to design the RIS configurations.  In the final stage, we insert additional
pilots to estimate the overall effective channel (or if the pilot budget is
limited, reuse the existing pilots), then re-optimize the beamformers.
%using the Weighted MMSE (WMMSE) algorithm\cite{wmmse} based on explicitly estimated channels.  The pilot overhead of the proposed framework is shown to be scalable to channel with varying channel coherence periods by adopting two types of pilot transmission schemes. 
%To do so, we could either 1) adopt an additional pilot training stage with the optimized RIS coefficients and scheduling fixed, so that the low-dimensional effective channels of scheduled users can be accurately estimated using very few pilots, or 2) re-use the received pilots in previous stages to estimate the high-dimensional channels, as channels are fixed within a channel coherence period. 
Numerical results show that the proposed algorithm can learn to maximize the network utility with significantly reduced pilot overhead as compared to the conventional channel estimation based approach, and can generalize to a network with an arbitrary number of users.

\subsection{Related Work}

User scheduling in RIS-assisted networks has not been thoroughly studied. Many existing proportional fairness based scheduling algorithms that have been successful in conventional networks without RIS\cite{PFscheduler1,PFscheduler2,PFscheduler3,PFscheduler4} are difficult to directly migrate to an RIS-assisted network, because the RIS 
can work in collaboration with BS beamformers to improve channel gain or suppress interference for some users, thus influencing scheduling decisions. 
Moreover, most of the published works involving RIS only focus on joint beamforming and reflection coefficients design, where the BS optimizes beamforming matrix and RIS phase shift vector based on perfect CSI to improve energy efficiency\cite{8wuqq,irs_perfcsi1,energy_effi}, to maximize weighted sum rate\cite{hguo,wsr_2,wsr_3}, to maximize the minimum rate\cite{2qua,irs_perfcsi2,minrate_3}, to maximize the system throughput\cite{aerial} or to maximize the received signal-to-interference-and-noise ratio (SINR)\cite{spectrumlearning}. None of these works address the problem of user scheduling in dense RIS-assisted networks. Also, most of these works are based on the perfect CSI assumption, which is an unrealistic assumption as the RIS, due to its passive structure, is not equipped with radio frequency components to perform active pilot symbol transmission, or signal processing units as receivers to estimate the channel\cite{bible}.

%Thus, many channel estimation protocols are recently proposed to re-evaluate the promising gain of RIS in a practical setting. \cite{channelest_diffmode} proposes a two-stage channel estimation scheme where the RIS is set to different mode to estimate different channels for each stage, to decompose the total channel estimation problem into a series of independent problems. Further in \cite{channelest_threestage}, the authors propose a three-phase pilot-based channel estimation framework that aims to reduce pilot overhead by exploiting the redundancy in the user-IRS-BS reflected channels. 

Recently, data-driven approaches have been proposed to address the challenge in
CSI acquisition and transmission optimization. This paper is inspired by the
successes in \cite{emil2020deep,taojournal, kareemdnn}, which show that it is
possible to bypass explicit channel estimation and to directly use the received 
pilot symbols to optimize the downlink beamforming configuration in their 
respective network setting. In an RIS-assisted network, \cite{emil2020deep} uses a deep
neural network with supervised learning to approximate the mapping from the
received pilot sequences to the optimized RIS phase matrix and downlink
beamforming vector for a single-user setting. In a multiuser setting,
\cite{taojournal} uses a GNN to exploit the permutation invariant and permutation
equivariant properties of multiuser
sum-rate maximization problem, and to learn the mapping from the received pilot
sequences to the RIS configuration and a set of beamforming vectors in an
unsupervised fashion. In the hybrid beamforming network, \cite{kareemdnn} utilizes
a deep neural network to design downlink analog beamformer based on the received
pilot frames, thus bypassing channel estimation.  Those works have shown that
the rich information in the received pilot symbols is sufficient to directly design
beamforming configuration. However, their proposed solutions do not account for
user scheduling, which is a crucial yet challenging component of network
design.

%weather the information is sufficient to design user scheduling is not studied. Those works have shown that a neural network can model the connection between received pilot symbols and beamforming configuration; however, the connection between pilot symbols to user scheduling is not explored. 

In \cite{Cui_2019_spatialdeep}, the authors perform link scheduling using a
novel deep neural network architecture based solely on the geographic locations of the devices.
Similarly, \cite{lee2020graph} makes use of location information and tackles link scheduling problem using a deep learning based graph embedding process. 
In \cite{zhao2021link}, the link scheduling problem is solved using a graph 
convolutional network based solution with user interference relationship as 
the primary input. However, these existing works on scheduling are for 
device-to-device networks without the RIS.  In this paper, we 
investigate the mapping from the received pilots to a user schedule and the
associated RIS and beamforming design, in recognition of the rich channel
information contained in the pilot symbols. Specifically, we adopt a
data-driven approach to learn such a mapping, and propose a deep learning based
multi-stage framework that uses the received pilot symbols as input to perform
joint scheduling and beamforming in an RIS-assisted network with reduced pilot
overhead.

%Then, we adopt data driven approach to learn the mapping and propose a deep learning based multi-resolution framework that mainly use received pilot symbols as input to performs joint scheduling and beamforming in RIS-assisted network with reduced pilot overhead%., in recognition of the rich channel information contained in the pilot symbols. 

\subsection{Main Contribution}

%This paper assumes that there exists an ideal mapping from received pilots to an user scheduling that is compatible with RIS and beamformer to achieve high network throughput while ensuring fairness. 
We propose to learn the mapping from the received pilot sequences to user scheduling, RIS configuration and beamformer using a GNN and train the model to directly maximize the network objective. A key motivation for adopting GNN as the scheduler is that it naturally captures the permutation invariant and equivariant properties of the optimization problem, i.e., regardless of the ordering of the users, it should result in the same scheduling with the same RIS reflection coefficients and the same beamforming vectors with permuted indices. GNN also allows generalizability across the users. 
For instance, the number of users in a dense network is constantly changing
depending on the user traffic. If we adopt a fully connected neural network as
the scheduler, the scheduler would not be able to generalize to different sizes
of user pool except by re-training. In contrast, a GNN can generalize to
different numbers of users by simply adding and removing components 
in its feature extraction and information
exchange stages \cite{PIandPE}.

We propose a permutation invariant and equivariant GNN architecture for mapping
the received pilots to the optimized variables. In particular, the RIS and the 
users are modelled as nodes in the GNN and each node is associated with a nodal
feature, also known as the representation vector. The input features, i.e.,
received pilots, are encoded into the representation vectors. Subsequently, the
representation vectors are updated layer-by-layer by exchanging information with
the neighboring nodes. After updating through multiple layers, the updated
representation vectors contain the right representation of information to design the RIS
configuration and beamformers, from which the schedule can be inferred. 

However, a large fully connected GNN is difficult to train with
high-dimensional features, because the functional landscape over a large number
of users with a nonconvex objective is highly non-trivial. To address this
scalability issue, we propose to use two GNNs. The first GNN takes very short 
received pilots and the user weights from all potential users as inputs in 
order to make the scheduling decisions in a first stage. We then use a second 
GNN to design the RIS configuration for the scheduled users in a second stage.
The second GNN works on a smaller set of users, so it can take longer pilot
sequences as nodal features. 

Using short pilot sequences to determine the schedule in the first stage is a
reasonable approach, because unlike beamforming and RIS configuration,
scheduling is not a strong function of the channel realizations. Thus, coarse
knowledge about the channel is already sufficient to enable the scheduler to
choose the (near) optimal set of users. After the schedule is determined, 
a second GNN that takes input from the scheduled users only, but with longer
pilots, are then used in the second stage to determine the RIS configurations.

% More specifically, the users can be scheduled based on the output beamforming vectors from the GNN scheduler, where users with significant allocated transmission power are selected. 

Further, %we find that the beamformers can be fine-tuned for additional performance gain. Therefore, 
we propose a beamformer fine-tuning stage as the
third stage to refine the beamformers using the weighted minimum mean square
error (WMMSE) algorithm \cite{wmmse} based on the explicitly estimated channels
for additional performance gain. 
%To ensure that the pilot overhead of the proposed framework is scalable to channels with different fading characteristics, 
Depending on the pilot overhead constraint, we adopt two choices of channel
estimation schemes. The first scheme involves an additional pilot training
with a fixed user schedule and fixed RIS coefficients. In this case, the
low-dimensional effective channels of the scheduled users can be accurately
estimated using additional short pilots in each scheduling instance. 
%Now, although the additional pilot training overhead to optimize beamformer at each scheduling instance is small, the pilot overhead can add up if there are a large number of scheduling slots. 
%Thus, this scheme is most suitable for fast-fading scenarios in which the number of scheduling instances is limited, or if the overall pilot constraint is not overly limiting. %, e.g., fast-fading channel, where the additional pilot overhead does not contribute much to the overall pilot training overhead. As a result, to adapt to 
But if the total pilot overhead budget is limited, 
we can also employ a second channel estimation scheme which 
%involves no additional pilot training to optimize BS beamformers in the third stage; the received
reuses the pilots in the previous stages to estimate the high-dimensional
channels, then subsequently design the beamformers based on 
the combined channels. The conference version of this paper \cite{zhongzeicassp} considers only the first scheme. This journal version of the paper considers both channel estimation schemes which enables the proposed framework to adapt to both fast and slow-fading environments.

In summary, the main contributions of this paper are as follows:
\begin{enumerate}
    \item Two separate GNNs are used first to schedule a subset of users, then to design the RIS reflection coefficients, so that together with the optimized BS beamformers, the users can achieve high throughput while ensuring fairness. Both GNNs use the received pilot symbols and the user priority weights as inputs. The model is permutation invariant with respect to the schedule and the RIS reflection coefficients, and permutation equivariant with respect to the beamforming vectors. It allows generalizability to different sizes of the user pool. 
    
    \item A hybrid data-driven and CSI-based framework is used, where the user schedule and the RIS reflection coefficients are designed using GNNs, and the beamformers are designed based on the estimated CSI. The framework contains three stages, where scheduling and RIS design are separated into the first and the second stages in order to reduce the neural network training complexity. A third stage is used to fine-tune the beamformers.
    
    %\begin{enumerate}
    %    \item In scheduling stage, a GNN model maps user weights and short pilot sequences of all users to beamforming vectors designed for all users. Subsequently, the schedule is inferred from the beamforming vectors by adopting the implicit scheduling scheme.
    %    \item In RIS design stage, a GNN model maps user weights and pilot sequences of scheduled users to RIS configuration and beamforming vectors designed for scheduled users.
    %    \item In beamformer fine-tuning stage, the combined channel of scheduled user is estimated to refine the beamformer using Weighted MMSE (WMMSE) algorithm.
    %\end{enumerate}
    
    \item An uplink pilot placement strategy is proposed, in which the pilot symbols received over random setting of the uplink RIS phases are used to design the user schedule and the RIS reflection coefficients, with possibly additional pilot symbols (received over the optimized uplink RIS configuration) used to design the BS beamformers. 

    \item Numerical simulations show that the proposed framework achieves better network utility compared to the conventional two-step approach, and can generalize to scenarios with an arbitrary number of users.
    
\end{enumerate}

\subsection{ Organization of the Paper and Notations}
%$\emph{Organization}$: 
The remaining paper is organized as follows. Section \ref{sec.background} introduces the system model, proportional fairness scheduling, and problem formulation. Section \ref{sec.gnn} describes the proposed pilot placement structure, deep learning framework and GNN architecture. Numerical results are provided in Section \ref{sec.numerical_irs}. The paper concludes with Section \ref{sec.conclusions}.

$\emph{Notations}$: We use $a$, $\bm{a}$, and $\bm{A}$ to denote scalar, vector, and matrix respectively; $\bm{A}^\top$, $\bm{A}^{\sf H}$ and $\bm{A}^{-1}$ to denote transpose, Hermitian and inverse; $|\cdot|$ and $(\cdot)^\ast$ to denote the modulus and conjugate; $[\bm{a}]_j$ to denote the $j$-th element of vector $\bm{a}$;
$\text{vec}(\bm{A})$ to denote matrix $\bm{A}$ in vector form; 
and diag($\bm{a}$) to denote the diagonal matrix with the entries of $\bm{a}$ on the diagonal. 
We use $\mathcal{R}(\cdot)$ and $\mathcal{I}(\cdot)$ to denote the real and imaginary component of a complex value; 
$\mathcal{C}\mathcal{N}(\cdot,\cdot)$ to denote a complex Gaussian distribution; $\mathbb{E}(\cdot)$ to denote the expectation of a random variable. 
Finally, we use $|\mathcal{S}|$ to denote the cardinality of the set $\mathcal{S}$, and $\bm{\text{I}}_b$ to denote a $b \times b$ identity matrix.

\section{System Model and Problem Formulation}
\label{sec.background}

\subsection{System Model}
\label{sysmodel}
\begin{figure}
  \includegraphics[width=\columnwidth]{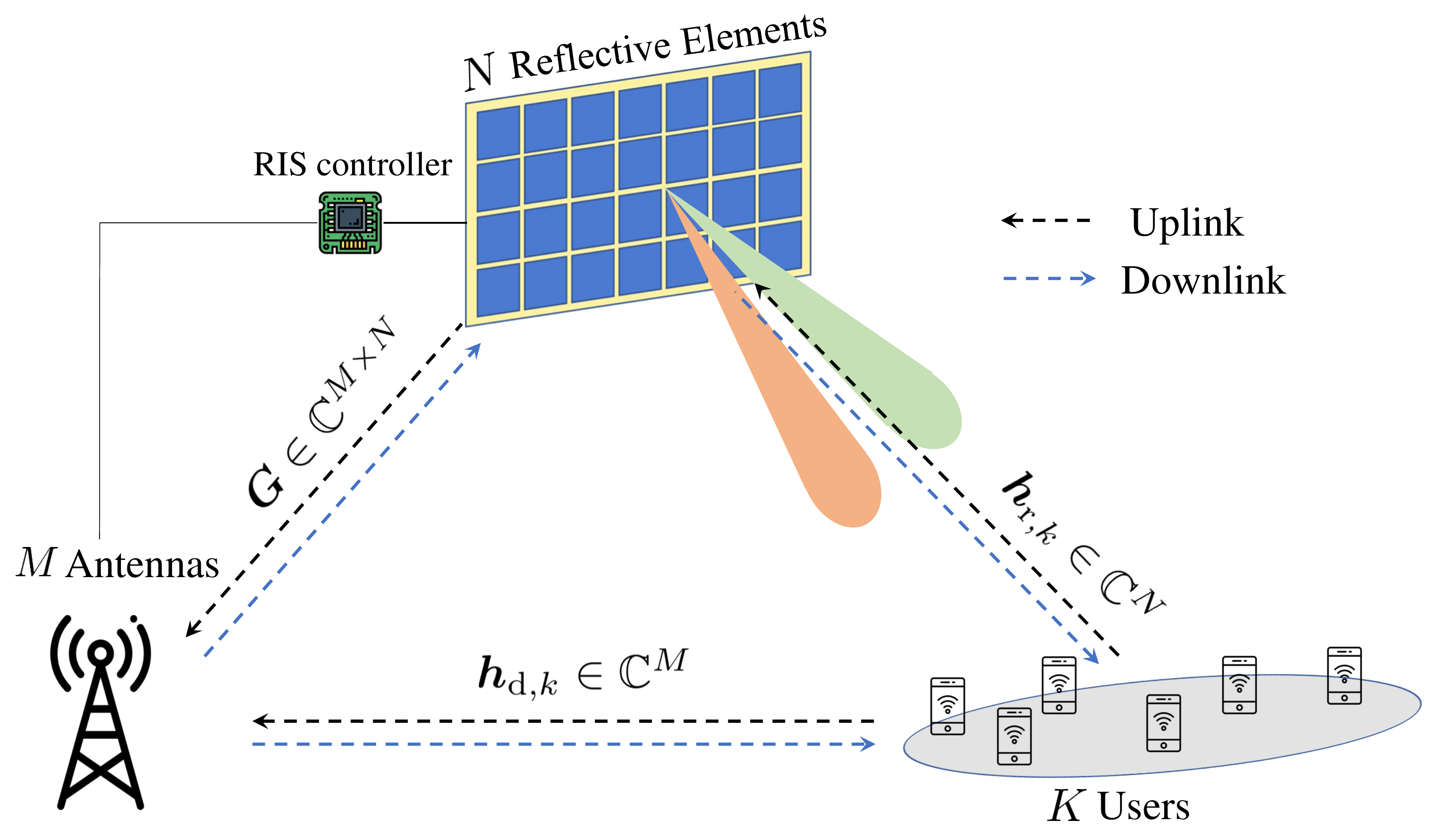}%
  \caption{RIS-assisted multiuser network}
  \label{fig.irs_background}%
\end{figure}

Consider a downlink RIS-assisted multiuser MISO network, in which a BS equipped with $M$ antennas serves $K$ single-antenna users as shown in Fig.~\ref{fig.irs_background}. 
An RIS with $N$ passive elements is placed between the BS and the users to enhance the SINR of the received signal at the users.
%, by making the signal in the reflection path and direct path add coherently or incoherently via an RIS controller. 
The RIS reflection coefficients are denoted as $\bm{\theta}=
[e^{j\delta_1}$,  $e^{j\delta_2}, \cdots , e^{j\delta_N}]^T\in\mathbb{C}^{N}$
with  $\delta_n \in [0,2\pi)$ as the phase shift of the $n$-th element.
These reflective coefficients can be controlled independently, from an RIS controller, typically located at the BS.

We consider the scenario in which the number of users in a cell is much larger
than the number of antennas at the BS i.e., $K > M$, whereas the BS can only
serve at most $M$ users simultaneously by employing spatial multiplexing.
Thus, the users need to take turns to get served.  The BS serves each set of
scheduled users in what is called a scheduling timeslot. The duration of the 
scheduling slots is determined by the system latency requirement, and is 
typically in the order of milliseconds.

To maximize the network utility function while ensuring fairness amongst users, the scheduler needs to strategically choose a subset of users in each scheduling timeslot. 
%It is crucial to schedule the users in each timeslot so that the network utility function of all the users can be maximized. 
Let $\beta_k^t \in \{0,1\}$ be the indicator variable that indicates the scheduling decision of the user $k$ at the $t$-th scheduling timeslot
\begin{equation}
\beta_k^t = 
    \begin{cases}
      0 & \text{if not scheduled,}\\
      1 & \text{otherwise.}
    \end{cases} 
\end{equation}
%if $U_k$ is scheduled, $\beta_k=1$; otherwise $\beta_k=0$, such that $\sum_{k=1}^{K}\beta_k \le M$,
Let $\bm{\beta}^t=[\beta_1^t,\beta_2^t,\cdots,\beta_K^t]^{\top}\in\mathbb{C}^{K}$. Then, we have $\|\bm{\beta}^t\|_1\le M$ since at most $M$ users are scheduled in a scheduling timeslot.

We adopt a \emph{block-fading} model in which the channels are assumed to be constant across multiple scheduling timeslots within a coherence period, then change independently in subsequent coherent periods. 
The number of scheduling timeslots within the channel coherence period depends
on the channel fading characteristics.  If the channel is fast-fading with a relatively short channel coherence period, there would be only a few scheduling
timeslots within the coherence period.
Conversely, if the channel is slow
fading, many more scheduling timeslots would be available within a channel
coherence period. 

The channel models are as shown in Fig.~\ref{fig.irs_background}, where $\bm{h}_{{\rm{d}},k}\in\mathbb{C}^{M}$ denotes the direct link channel from the BS to the user $k$, $\bm{h}_{{\rm{r}},k}\in\mathbb{C}^{N}$ denotes the reflected link channel from the RIS to the user $k$, and $\bm{G} \in\mathbb{C}^{M\times N}$ denotes the common channel from the BS to the RIS. 

Let $s_k^t\in\mathbb{C}$ be the information symbol to be transmitted from the BS to the user $k$ and let $\bm{w}_k^t\in\mathbb{C}^{M}$ denote the associated beamforming vector of the user $k$ at the $t$-th scheduling timeslot. We can denote the beamforming matrix at the BS as $\bm{W}^t = [\bm{w}_1^t, \cdots, \bm{w}_K^t] \in\mathbb{C}^{M\times K}$. In each scheduling timeslot, we have $\|\bm w_k^t\|^2=0$ if $\beta_k^t=0$, otherwise $\|\bm w_k^t\|^2>0$. The transmitted signal at the BS at the $t$-th scheduling timeslot can be denoted as:
%be the transmitting antenna at BS associated with $U_k$, where $\|\bm w_k\|^2 = 0$ if $\beta_k=0$, otherwise $\|\bm w_k\|^2 \neq 0$. The transmitted signal at the BS can be expressed as:
\begin{align}
&\centering
\bm{s}^t = \sum_{k=1}^{K}\bm{w}_k^t\beta_k^t s_k^t.
\end{align}

Consider a single $t$-th scheduling timeslot. The received signal at the user $k$ is a combination of the signals from the direct path and the reflection path through the RIS, which is given by
\begin{equation} 
\begin{split}
r_k^t & = \bm{h}_{{\rm{d}},k}^{\top}\bm{s}^t +  \left(\bm{G} \text{diag}(\bm{\theta}^t)\bm{h}_{{\rm{r}},k}\right)^{\top}\bm{s}^t + n_k\\
    & = \left(\bm{h}_{{\rm{d}},k}+\bm{A}_k\bm{\theta}^t\right) ^{\top}\sum_{i=1}^{K}\bm{w}_i^t\beta_i^ts_i + n_k,
\end{split}
\end{equation}
\noindent where $\bm{A}_k = \bm{G}\text{diag}(\bm{h}_{{\rm{r}},k})\in\mathbb{C}^{M\times N}$ is the cascade channel between the BS and the user $k$ through the reflection at the RIS, $\bm{\theta}^t$ is the RIS reflection coefficients at the $t$-th scheduling timeslot, and $n_k \sim\mathcal{C}\mathcal{N}(0,\sigma_d^2)$ is the downlink additive white Gaussian noise. 

The achievable rate of the user $k$ at the $t$-th scheduling timeslot can be expressed as:
\begin{align}\label{realrk}
&\centering
R_k^t = \log \left(1 + \dfrac{|(\bm{h}_{{\rm{d}},k}+\bm{A}_k\bm{\theta}^t)^{\top}\bm{w}_k^t|^2\beta_k^t}{\sum_{i\neq k} |(\bm{h}_{{\rm{d}},k}+\bm{A}_k\bm{\theta}^t)^{\top}\bm{w}_i^t|^2\beta_i^t + \sigma_d^2}\right).
\end{align}
Here, we assume that the effective channel gains can be accurately estimated at the user side using a few additional downlink pilots prior to data transmission, so the rate in (\ref{realrk}) is achievable\cite{modelbased}.

\subsection{Proportional Fairness Scheduling}

%Proportional fairness scheduling is the core 
To ensure fairness in user scheduling, we define a network utility, which is a function of the long-term average rate achieved by each user, computed according to 
an exponentially weighted moving average: 
\begin{align}
&\centering
\bar{R}_k^t = (1-\gamma)\bar{R}_k^{t-1} + \gamma R_k^{t-1},
\end{align}
\noindent where $0\le \gamma \le 1$ is the forgetting factor and $\bar{R}_k^t$ is average rate of the $k$-th user prior to the $t$-th scheduling timeslot. Defining a non-decreasing and concave utility function $U( \cdot )$ for each user, the network-utility maximization problem is that of maximizing 
\begin{align} \label{sum_utility}
\sum_{k=1}^{K} U(\bar{R}_k^t).
\end{align}

In the proportional fairness scheduling problem, the utility function is chosen as $U(\cdot) = \log(\cdot)$. Assuming a small $\gamma R_k^{t-1}$, the maximization of log-utility can be approximated by the maximization of weighted sum-rate \cite{logutility_to_wsr}, so that problem (\ref{sum_utility}) can be approximately solved by maximizing the following:
\begin{align}\label{wsr}
&\centering
\sum_{k=1}^{K} \alpha_k^t R_k^t,
\end{align}
\noindent where the weight $\alpha_k^t$ is the inverse of $\bar{R}_k^t $:
\begin{equation} \label{weightupdate}
\begin{split}
\alpha_k^t & =  \left.\frac{\partial U(\bar{R}_k^t)}{\partial R} \right|_{R=\bar{R}_k^t} =  \left.\frac{\partial \text{log}(\bar{R}_k^t)}{\partial R} \right|_{R=\bar{R}_k^t} =\frac{1}{\bar{R}_k^t}.
\end{split}
\end{equation}

%\color{blue}
%The original network objective (\ref{sum_utility}) can be approximately solved by a series of weighted sum-rate maximization problem, where the weights $\alpha_k^t$ are updated at each timeslot according to (\ref{weightupdate}). 
%In next section, we present the problem formulation of proportional fairness scheduling based on (\ref{sum_utility}).

%\color{black}
%Therefore, we can solve the proportional fairness scheduling problem in long term by solving a weighted sum-rate maximization problem at each scheduling timeslot. 

\subsection{Problem Formulation}
To maximize the weighted sum rate in \eqref{wsr}, it is necessary to acquire the knowledge about the channels. Assuming channel reciprocity and that the system operates in the time division duplex (TDD) model, we rely on an uplink pilot transmission phase to gain information about the channel. 
The channel coherence period is therefore divided into two phases, i.e., Phase-\RN{1} for uplink pilot training and Phase-\RN{2} for downlink data transmission.
Downlink data transmission phase consists of multiple scheduling timeslots as discussed in Section \ref{sysmodel}. 

In the pilot training phase, all $K$ users transmit pilot sequences at the same time. Let the pilot sequence of the $k$-th user be $x_k(\ell)$, $\ell = 1,\cdots,L$. The BS receives a combination of the signal from the direct path and the signal reflected off the RIS, so the received pilots at the BS can be expressed as
\begin{align}
    \bm y(\ell)= \sum_{k=1}^K(\bm h_{{\rm{d}},k}+\bm A_k\bm \theta(\ell)) x_k(\ell)+\bm n(\ell), \ell = 1,\cdots, L,
\end{align}
\noindent where $\bm{\theta}(\ell)$ is the uplink RIS configuration at the $\ell$-th instance and $\bm{n}(\ell) \sim\mathcal{C}\mathcal{N}(0,\sigma_u^2\bm{I})$ is the uplink additive noise vector.

In this paper, we focus on optimizing the system variables at each timeslot based on instantaneous CSI (in contrast to works that use statistical CSI, e.g.,~\cite{van2021controlling}). The conventional system design would first estimate the channels then optimize the system variables.  A key observation here is that since the ultimate goal is to maximize the network utility (\ref{sum_utility}), %This paper proposes to exploit the pilots more efficiently by mapping the received pilot sequences to the scheduling decision, RIS configuration and beamforming variables.
instead of explicitly estimating the CSI, we can pursue a data-driven approach to the system-level optimization problem by directly mapping the received pilots and the user weights to the optimized user schedule, RIS configuration, and the beamforming
variables. 

Conceptually, the optimization problem in each scheduling slot can
be thought of as:
%More specifically, we propose to design scheduling decision $\bm{\beta}^t$, beamforming matrix $\bm{W}^t$ and RIS reflection coefficients $\bm{\theta}^t$, directly based on received pilot sequences $\bm{Y}$ and user weights $\bm{\alpha}^t$. 
%As shown in Fig.\ref{fig.pilotplacement}, the mapping from input to optimal scheduling/beamforming is to be done repeatedly within each channel coherent time block, as user weight is a function of past instantaneous rates as it updates from timeslot to timeslot. 
%Under a downlink transmission power constraint $P_d$, the joint scheduling and beamforming problem can be expressed as follows:
\begin{equation} \label{bigobjective}
\begin{aligned}
\underset{\scriptsize \begin{array}{ll} (\bm \beta^t ,\bm W^t, \bm \theta^t) = \\ f(\{\bm{y}(\ell)\}_{\ell=1}^{L}, \bm \alpha^t)\end{array} }{\;\;\;\;\;\textrm{maximize}\;\;} & \sum_{k} \alpha_k^t R_k^t\\
\textrm{subject to} \quad &\sum_{k} \|\bm{w}_k^t\|^2 \le P_d,~
\lvert{[\bm{\theta}^t]_n}\rvert= 1, ~\forall n,\\
  &\sum_{k}\beta_k^t \le M, \beta_k^t \in \{0,1\}, ~\forall k,
\end{aligned}
\end{equation}
where $\bm \alpha_t=[\alpha^t_1,\cdots,\alpha^t_K]^\top$ and $P_d$ denotes the downlink transmission power constraint.

Finding the optimal functional mapping $f(\cdot)$ for problem (\ref{bigobjective}) is computationally challenging as it is a mixed discrete (scheduling) and continuous (RIS reflection coefficients,  beamforming matrix) optimization problem with nonconvex objective and nonconvex constraints. 
%The scheduling subproblem alone is a nonconvex integer programming problem, and the beamforming subproblem is a variational optimization problem also with a nonconvex objective function. 
Moreover, it is difficult to decouple the optimization variables as they are closely interrelated. For example, to minimize interference, we should schedule users whose channels are orthogonal, i.e., users who are geographically far from one another, but in order to leverage the full benefit of RIS, the scheduled users should be in close proximity in order to take the advantage of the more directional beams from the RIS. 
This type of trade-off calls for a solution that is capable of modelling the dependencies between the optimization variables. 

To address the challenges in solving problem (\ref{bigobjective}), we propose to utilize deep neural network as a powerful function approximator\cite{functionapprox} to model the mapping function $f(\cdot)$ and learn the transmission strategy from data.
However, designing a single neural network to learn such a mapping is not trivial, due to its complicated functional landscape. 
In the next section, we describe a multi-stage approach to learn such a mapping. %The pilot placement structures that enables the algorithm to adapt to both fast-fading channel and slow-fading channel is first introduced. 

\section{Multi-Stage Learning Framework}
\label{sec.gnn}

Weighted sum rate can be a challenging objective to learn. It has been shown in \cite{Cui_2019_spatialdeep} that the addition of user weights as input to the neural network imposes new learning challenge.
This is due to the fact that a small change in user weights can drastically change the scheduling and the associated RIS design and beamforming vectors. Thus, to learn the function mapping, the neural network would need to sample a large amount of data, including the entire space of user weights, but the distributions of user weights are highly non-uniform and difficult to sample. 

%Here, we make a key observation that only need very few
%pilots to estimate the combined channel $\bm{h}_{{\rm},k}+\bm{A}_k\bm{\theta}^t$.  
In this paper, we propose a multi-stage approach to solving the weighted
sum-rate maximization problem (\ref{bigobjective}). The idea is that user
scheduling and RIS configuration can be designed in the first two stages in a
data-driven fashion with one set of uplink pilots. Once the scheduled users and
the RIS reflection coefficients are fixed, the optimal beamforming matrix is
now a function of the effective channel, and can be designed in a
third stage based on the estimated CSI, by either re-using the existing pilots
or with possibly additional pilots.

%relatively very few pilots. Subsequently, for high efficiency in pilot processing, then design the beamformer based on the estimated combined channel.  For \emph{1}) scheduling, RIS reflection and \emph{2}) beamforming, we propose two different pilot training schemes. 

Below we first describe the proposed uplink pilot placement strategy, 
then the GNN architecture for learning the user schedule in stage one and 
the RIS configuration in stage two, and finally the design of beamformers 
in stage three.

\subsection{Pilot Placement for Scheduling and RIS Design}\label{pilot_training}

\begin{figure}[!t]
\centering
\begin{subfigure}{\columnwidth}
	\centering
\includegraphics[width=\columnwidth]{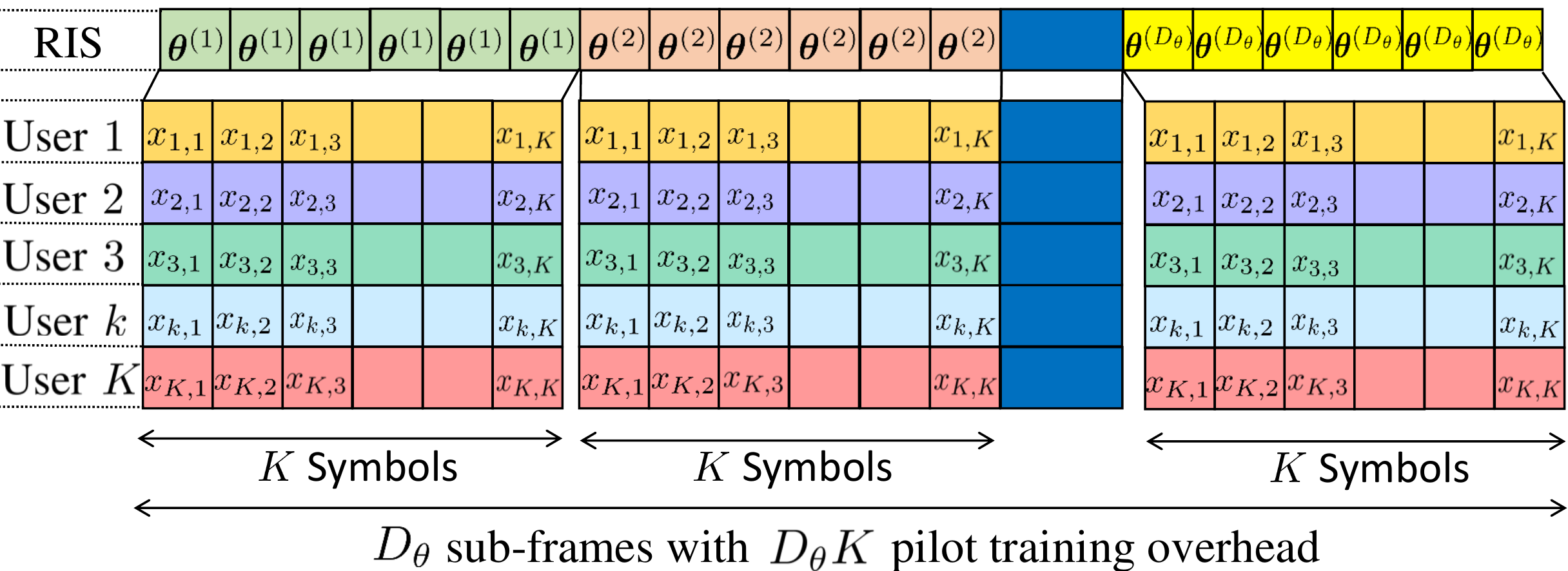}
\caption{$D_{\theta}$ pilot sub-frames from $K$ users, over random uplink RIS configuration}
\label{fig.pilotframe_dtheta}
\end{subfigure} \\

\vspace{1em}
\begin{subfigure}{\columnwidth}
	\centering
\includegraphics[width=\columnwidth]{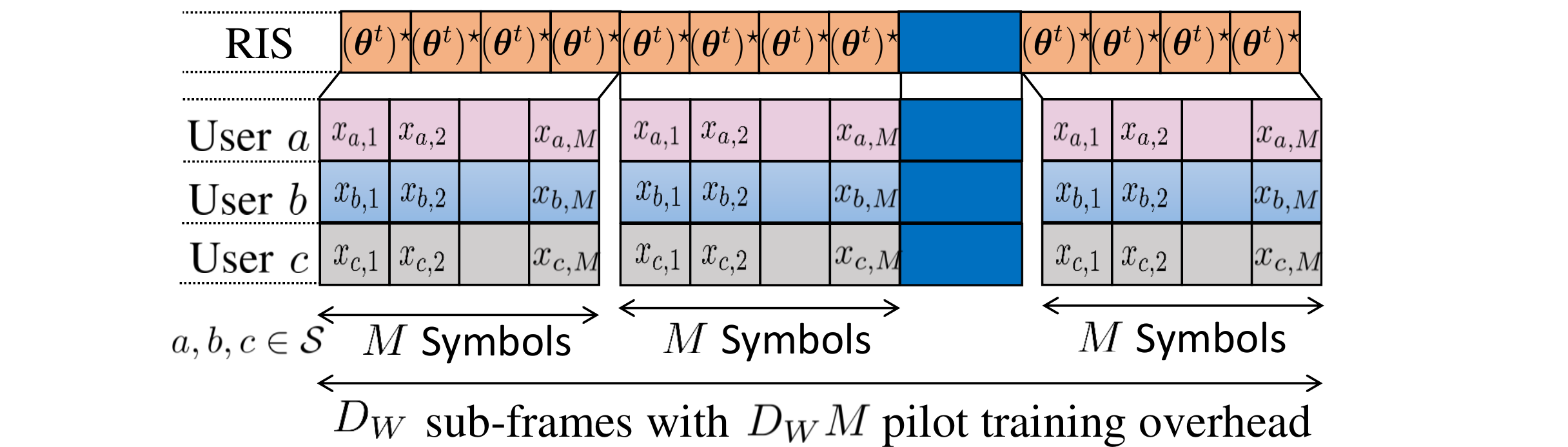}%
\caption{$D_{W}$ pilot sub-frames from scheduled users, over optimized uplink RIS configuration}%
\label{fig.pilotframe_dw}%
\end{subfigure}
\caption{Uplink pilot frame structure for: (a) scheduling and RIS configuration; 
and (b) beamforming.}
\label{fig.pilotframe}
\end{figure}

\begin{figure*}[t]
\begin{subfigure}{\linewidth}
\centering
\includegraphics[width=0.53\columnwidth]{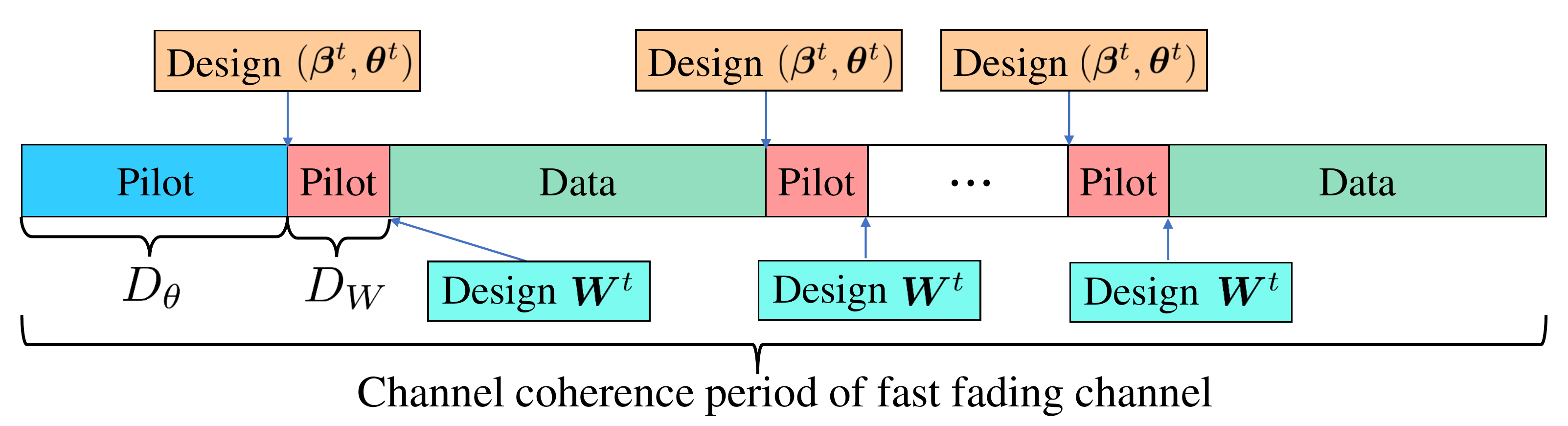}%
\caption{Design beamformers by estimating the combined channels using additional $D_W$ pilots per scheduling slot.}%
\label{fig.pilotplacement_fast}%
\end{subfigure} \\

\vspace{1em}
\begin{subfigure}{\linewidth}
\centering
\includegraphics[width=0.82\columnwidth]{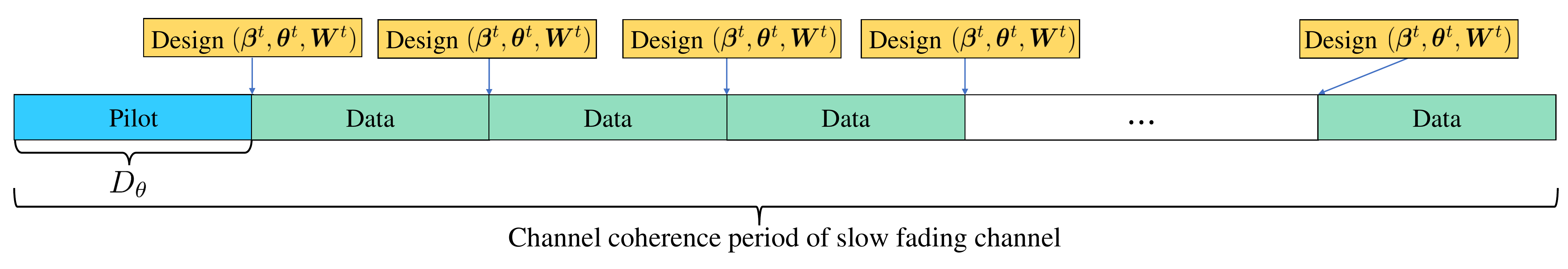}%
\caption{Design beamformers based on estimating the high-dimensional channel.}%
\label{fig.pilotplacement_slow}%
\end{subfigure}
\caption{Pilot placement structures for scheduling, RIS configuration and beamformer design over a coherence period.}
\label{fig.pilotplacement}
\end{figure*}

For scheduling and RIS design in the first two stages, we adopt the pilot transmission protocol
proposed in \cite{jiechannelest} for pilots design and the uplink 
reflection coefficients at the RIS in the uplink pilot phase. 
The user pilots are set to be orthogonal, denoted as $\bm{x}_k$'s, where
$\bm{x}_k^{\sf H} = \left[x_{k,1},x_{k,2},\cdots,x_{k,K}\right]$. 
Note that since there are a total of $K$ potential users in the scheduling pool, the user pilot length is set to be $K$ to ensure pilot orthogonality. 
%The proposed protocol leverages the orthogonality of the pilot sequences and sufficient observations of different RIS phase shift pattern. That is, 
The total training phase is equally partitioned to $D_{\theta}$ sub-frames, 
where each sub-frame is of symbol duration $K$, as shown in 
Fig.~\ref{fig.pilotframe}(\subref{fig.pilotframe_dtheta}). 
The reflection coefficients at the RIS are randomly chosen from sub-frame to
sub-frame but remain fixed within each sub-frame, while the users repeatedly
transmit the same mutually orthogonal pilot sequences over the
$D_\theta$ sub-frames. Let $\bm{\theta}^{(d)}$ be the
uplink RIS configuration in the $d$-th sub-frame.  The overall received pilots
in the $d$-th sub-frame $\bm{Y}^{(d)}\in\mathbb{C}^{M\times K}$ are given by
\begin{equation} 
\begin{aligned}
\bm{Y}^{(d)} = \sum_{k=1}^{K} \left(\bm{h}_{{\rm d},k}+\bm{A}_k\bm{\theta}^{(d)}\right)\bm{x}_k^{\sf H} + \bm{N}^{(d)}, d = 1, \cdots, D_{\theta},
\end{aligned}
\end{equation}
\noindent where $\bm{N}^{(d)}$ is the noise matrix whose columns are independently and identically distributed following the distribution $\mathcal{C}\mathcal{N}(0,\sigma_u^2\bm{I})$. We can leverage the orthogonality of the received pilots sequence in each sub-frame and decorrelate $\bm{Y}^{(d)}$ at the BS. Specifically, since $\bm{x}_k^{\sf H}\bm{x}_i = 0$ if $i\neq k$ and $\bm{x}_k^{\sf H}\bm{x}_k = KP_u$, with $P_u$ as the uplink pilot transmission power, the contribution from the user $k$ in the $d$-th sub-frame is given by:
\begin{equation}
\begin{split}
\bm{y}_{k}^{(d)} & = \dfrac{1}{K}\bm{Y}^{(d)}\bm{x}_k.%\\
%&=\bm{h}_{{\rm d},k}+\bm{A}_k\bm{\theta}^{(d)}+\bm{ n}_k^{(d)} \\
%& \triangleq \bm{H}_k\bm{q}^{(d)} + \bm{ n}_k^{(d)},
\end{split}
\end{equation}
\noindent %where $\bm{ n}_k^{(d)}\triangleq \bm{N}^{(d)}\bm{x}_k/K$. We denote the combined channel matrix as $\bm{H}_k \triangleq [\bm{h}_{{\rm d},k},\bm{A}_k]\in\mathbb{C}^{M\times (N+1)}$ and denote the combined phase shifts as $\bm{q}^{(d)} \triangleq [1,{\bm{\theta}^{(d)}}^{\top}]^{\top}\in\mathbb{C}^{(N+1)}$. 
%Given that there are $D$ sub-frames in total, BS receives $D$ copies of orthogonal pilot sequences of length $K$ from the $k$-th user. 
The collection of received pilot sequences from the user $k$ in overall $D_{\theta}$ sub-frames can be denoted as:
\begin{equation} 
\begin{split}
\bm{{Y}}_k^{D_{\theta}} &= \left[\bm{{y}}_{k}^{(1)}, \bm{{y}}_{k}^{(2)},\cdots,\bm{{y}}_{k}^{(D_{\theta})}\right].%\\
%&=\bm{H}_k\bm{Q}+\bm{ N}_k,
\end{split}
\end{equation}
%\noindent %where $\bm{ N}_k=[\bm{ n}_k^{(1)},\cdots,\bm{ n}_k^{(D)}]\in\mathbb{C}^{M\times D}$ and $\bm{Q}=[\bm{q}^{(1)},\cdots,\bm{q}^{(D)}]\in\mathbb{C}^{(N+1)\times D}$. 
Note that the overall pilot training overhead over a total of $D_{\theta}$ sub-frames is $D_{\theta}K$. 
This same set of received pilots $\bm{{Y}}_k^{D_{\theta}}$ are used to optimize
the user scheduling and the RIS configuration in the first two stages.

%In Section \ref{subsec:three-stage-framework}, we present a data-driven approach to optimize the scheduling and the RIS configuration based on 

%The $\bm{{Y}}_k^{D_{\theta}}$'s can be used as input to neural network to design system parameters\cite{taojournal}, or used to recover the unknown channel coefficients\cite{jiechannelest}. 

\subsection{Pilot Placement for Beamforming Design} 
\label{sec:pilot_beamformer}

For the beamforming design for the scheduled users in the third stage, two pilot strategies
are possible. Observe that once the scheduling and the RIS configuration are
fixed, the combined channel is now of lower dimension, and is given by
\begin{equation}
\bm{h}_{{\rm c},k} = \bm{h}_{{\rm d},k} + \bm{A}_k(\bm{\theta}^t)^{\star}, %~k\in \mathcal{S},
\end{equation}
where $(\bm{\theta}^t)^{\star}$ denotes the optimized RIS reflection coefficients
at the $t$-th scheduling timeslot. To obtain the combined channel $\bm{h}_{{\rm c},k}$, we have the option
of either: (i) estimating the low-dimensional channel with additional pilots training; 
or (ii) estimating the high-dimensional channel $\bm h_{{\rm{d}},k}$ and $\bm A_k$ 
without additional pilots training, then compute $\bm{h}_{{\rm c},k}$.
These two channel estimation strategies each have their advantages depending on
the total pilot overhead budget, and whether the channel is fast-fading or slow-fading.

\subsubsection{Estimating Low-Dimensional Combined Channel}

With fixed scheduling and RIS configuration, estimating the combined channel needs only relatively short pilots\cite{aggregatechannel}. 
%as compared to estimating the channel $\bm{A}_k$ and $\bm{h}_{{\rm d},k}$ separately. %Only few pilots are sufficient to accurately estimate the low-dimensional channel, from which we can design the beamforming matrix to a very satisfactory degree. 
In particular, since at most $M$ users are scheduled, we can assign orthogonal
pilots of length $M$ to the scheduled users, denoted as $\bm{\tilde{x}}_k$, 
over the optimized RIS reflection coefficient, in a separate uplink training 
phase, over $D_W$ sub-frames, as shown in Fig.~\ref{fig.pilotframe}(\subref{fig.pilotframe_dw}). 

The received pilots at the BS in the $d$-th sub-frame are given by
\begin{align}
    \bm{\tilde Y}^{(d)}= \sum_{k\in \mathcal{S}}\bm h_{{\rm c},k}\bm{\tilde x}_k^{\sf H}+\bm{\tilde N}^{(d)}, ~d = 1,\cdots, D_W,
\end{align}
where $\mathcal{S}$ denotes a set of scheduled user. Based on the orthogonality of the transmitted pilots, we decorrelate the received pilots to obtain the contribution from the $k$-th user:
    \begin{equation}
    \begin{split}
    \bm{\tilde{{y}}}_k^{(d)} & = \dfrac{1}{M}\bm{\tilde Y}^{(d)}\bm{\tilde x}_k\\
        %&=\bm{h}_{{\rm d},k}+\bm{A}_k(\bm{\theta}^{t})^{\star}+\bm{\tilde{{n}}}_k^{(d)}\\
        &\triangleq \bm{h}_{{\rm c},k} + \bm{\tilde{{n}}}_k^{(d)},
    \end{split}
    \end{equation}
where $\bm{\tilde{{n}}}_k^{(d)}\triangleq \bm{\tilde{N}}^{(d)}\bm{\tilde x}_k/{M}$. %We denote the combined channel matrix as $\bm{H}_{{\rm c},i} \triangleq [\bm{h}_{{\rm d},i},\bm{A}_i]\in\mathbb{C}^{M \times (N+1)}$ and denote the combined phase shifts as $\bm{q}^{(d)} \triangleq [1,{\bm{\theta}^{(d)}}^H]^H\in\mathbb{C}^{(N+1)}$. 
%Given that there are $D_W$ frames in total, BS receives $D_W$ copies of orthogonal pilot sequences of length $M$ from the $k$-th user. 
The collection of received pilot sequences from the user $k$ in overall $D_W$ sub-frames can be denoted as:
\begin{equation}
\begin{split}
\bm{\tilde{{Y}}}_k^{D_W} &= \left[\bm{\tilde{{y}}}_{k}^{(1)}, \bm{\tilde{{y}}}_{k}^{(2)},\cdots,\bm{\tilde{{y}}}_{k}^{(D_W)}\right].
%&=\bm{h}_{{\rm c},i}\mathbf{1}_{D_W}+\bm{\tilde{{N}}}_i
%& = \bm{{H}}_{{\rm c},i}+\bm{\tilde{{N}}}_i,
\end{split}
\end{equation}
We can then use linear minimum mean-squared error (LMMSE) estimation to estimate the combined channel of the $k$-th user:
\begin{IEEEeqnarray}{rCl}
{\bm{h}_{{\rm c},k}} & = & \mathbb{E}[\bm{h}_{{\rm c},k}]+(\bm {\tilde Y}_k^{D_{W}}- \mathbb{E}[\bm {\tilde Y}_k^{D_{W}}])\notag\\
    & & \quad \left(\mathbb{E}[(\bm {\tilde Y}_k^{D_{W}}-\mathbb{E}[\bm {\tilde Y}_k^{D_{W}}])^{\sf H}(\bm {\tilde Y}_k^{D_{W}}-\mathbb{E}[\bm {\tilde Y}_k^{D_{W}}])] \right) ^{-1}\notag\\
    & & \qquad \mathbb{E}[(\bm {\tilde Y}_k^{D_{W}}-\mathbb{E}[\bm {\tilde Y}_k^{D_{W}}])^{\sf H} (\bm{h}_{{\rm c},k}-\mathbb{E}[\bm{h}_{{\rm c},k}])].
\end{IEEEeqnarray}

\subsubsection{Estimating High-Dimensional Channels}

The pilot overhead of directly estimating a low-dimensional channel is
$M D_W$. This is to be performed at every scheduling timeslot, 
as shown in Fig.~\ref{fig.pilotplacement}(\subref{fig.pilotplacement_fast}), 
because the
effective low-dimensional channel is a function of the scheduled users and
the optimized RIS configuration, which are different in each scheduling timeslot.
%where scheduling and RIS configuration are first designed based on $\footnotesize \begin{array}{ll} \{\bm Y^{(d)}\}_{d=1}^{D_{\theta}}\end{array}$, then beamformer is designed based on the estimated channel with $\footnotesize \begin{array}{ll} \{\bm{\tilde Y}^{(d)}\}_{d=1}^{D_{W}}\end{array}$. Let $\Upsilon$ denote the total number of scheduling timeslots in the channel coherence period. The total number of pilot overhead over a channel coherence period is 
%\begin{equation}\label{overheadcalculation}
%L=K \times D_{\theta} + n(\mathcal{S}) \times D_W \times \Upsilon,
%\end{equation}
%As shown in equation \eqref{overheadcalculation}, 
%This scheme is most suitable in the fast-fading channel, where each channel coherence period contains only a few scheduling timeslot.
%channel coherence period is short, i.e., fast-fading channel, as the total pilot overhead scales with the number of scheduling timeslots. 
%To adapt to slow-fading channel with prolonged channel coherence period, 
%we maintain the multi-staged structure, where scheduling and RIS are first designed based on $\footnotesize \begin{array}{ll} \{\bm Y^{(d)}\}_{d=1}^{D_{\theta}}\end{array}$, but we optimized the beamformer by estimating high-dimensional channel $\bm{h}_{{\rm d},k}$ and $\bm{A}_k$ using the same pilots for scheduling and RIS design without additional pilot phase.
Although the above pilot overhead is small, 
when the channel is slow-fading so there are 
many scheduling timeslots within a channel coherence period, 
the total pilot overhead may still be considerable.
Thus, when the total pilot budget is limited, 
it may be more advantageous to re-use the received pilots $\{\bm Y^{(d)}\}_{d=1}^{D_{\theta}}$ to estimate the high-dimensional channel $\bm{h}_{{\rm d},k}$ and $\bm{A}_k$. In this way, the combined channel $\bm{h}_{{\rm c},k}$ can be obtained from $(\bm{h}_{{\rm d},k}, \bm{A}_k, (\bm{\theta}^t)^{\star})$ thus avoiding the extra pilot overhead, %In effect, the solution relies on one type of pilots: $\footnotesize \begin{array}{ll} \{\bm Y^{(d)}\}_{d=1}^{D_{\theta}}\end{array}$. 
as shown in Fig.~\ref{fig.pilotplacement}(\subref{fig.pilotplacement_slow}). 

Recall that the collection of the received pilots over $D_{\theta}$ from the $k$-th user for scheduling is $\bm{{Y}}_k^{D_{\theta}}$, whose column can be written as %The entries of $\bm{{Y}}_k^{D_{\theta}}$ are $\bm{{y}}_{k}^{(d)}$'s and contain information about the high-dimensional channel
\begin{equation}\label{channel_highdim}
\begin{split}
\bm{y}_{k}^{(d)} & =\bm{h}_{{\rm d},k}+\bm{A}_k\bm{\theta}^{(d)}+\bm{ n}_k^{(d)} \\
& \triangleq \bm{H}_k\bm{q}^{(d)} + \bm{ n}_k^{(d)},
\end{split}
\end{equation}
\noindent where $\bm{ n}_k^{(d)}\triangleq \bm{N}^{(d)}\bm{x}_k/K$. We denote the high-dimensional channel matrix as $\bm{H}_k \triangleq [\bm{h}_{{\rm d},k},\bm{A}_k]\in\mathbb{C}^{M\times (N+1)}$ and denote the combined phase shifts as $\bm{q}^{(d)} \triangleq [1,{\bm{\theta}^{(d)}}^{\top}]^{\top}\in\mathbb{C}^{(N+1)}$. We can estimate the channel matrix $\bm{H}_k$ using LMMSE estimation as follows:
\begin{IEEEeqnarray}{rCl}\label{lmmse}
    {\bm{ H}_k} & = & \mathbb{E}[\bm H_k] + (\bm { Y}_k^{D_{\theta}}- \mathbb{E}[\bm { Y}_k^{D_{\theta}}])\notag\\
    & & \quad \left(\mathbb{E}[(\bm { Y}_k^{D_{\theta}}-\mathbb{E}[\bm { Y}_k^{D_{\theta}}])^{\sf H}(\bm { Y}_k^{D_{\theta}}-\mathbb{E}[\bm { Y}_k^{D_{\theta}}])] \right) ^{-1}\notag\\
    & & \qquad \mathbb{E}[(\bm { Y}_k^{D_{\theta}}-\mathbb{E}[\bm { Y}_k^{D_{\theta}}])^{\sf H} (\bm H_k-\mathbb{E}[\bm H_k])].
    \end{IEEEeqnarray}
We can adjust $D_\theta$ to ensure the 
estimation accuracy of $\bm{h}_{{\rm d},k}$ and $\bm{A}_k$. Importantly,
no additional pilot training is needed for designing the beamformers, as shown in Fig.~\ref{fig.pilotplacement}(\subref{fig.pilotplacement_slow}).

\subsection{GNN for User Scheduling and RIS Configuration}

\begin{figure}[!t]
  \includegraphics[width=\columnwidth]{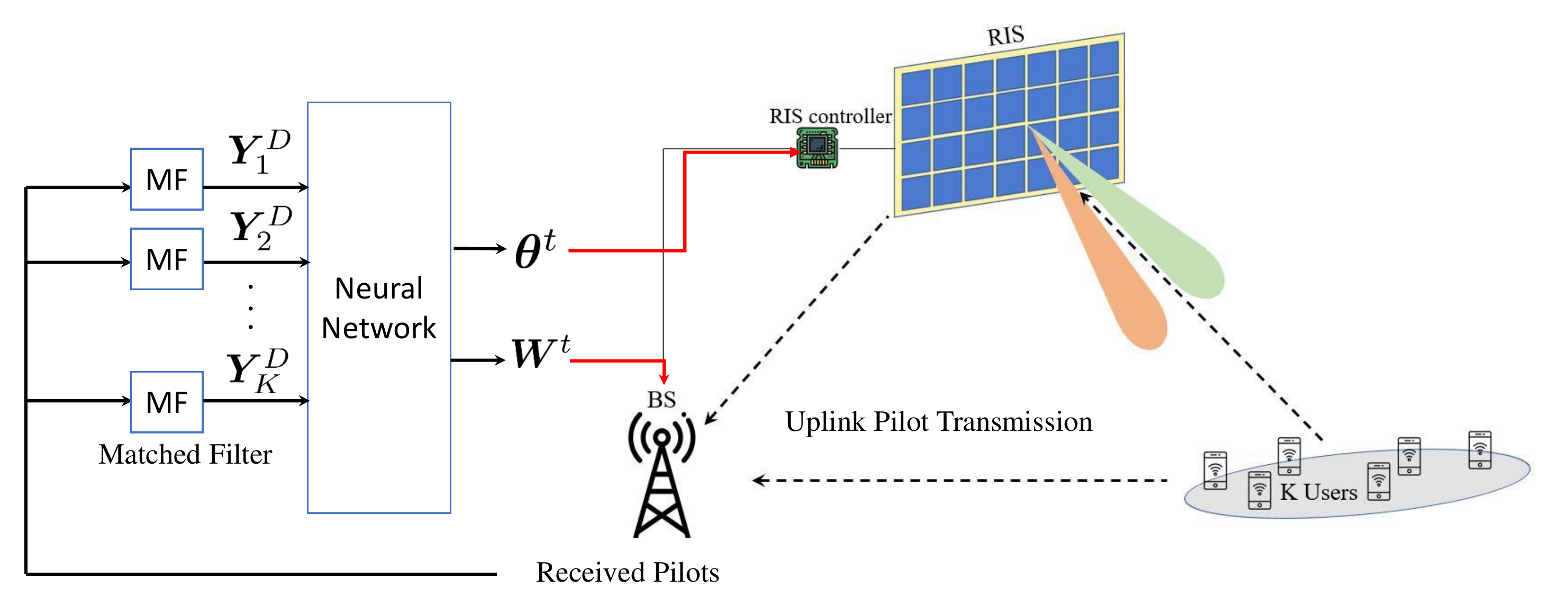}%
  \caption{GNN framework.}
  \label{fig.nodeconnection}%
\end{figure}

We now introduce the deep learning model for user scheduling and RIS configuration in the first two stages of the overall framework. %that can learn to optimize the user scheduling and RIS configuration from the receiver pilots $\{\bm Y^{(d)}\}_{d=1}^{D_{\theta}}$ and the user priority weights as the inputs, 
A key building block of the proposed data-driven approach is a GNN architecture
that takes the received pilots and the user weights as inputs and produces the
optimized RIS configuration and BS beamformers as output, as shown in Fig.~\ref{fig.nodeconnection}.

The use of GNN to
model the inter-relationship between the RIS and the users is crucial, because
the proposed GNN architecture allows certain permutation invariant and equivariant
properties to be observed. That is, if the ordering of the users is permuted, the
neural network should permute the set of beamforming vectors accordingly, while
keeping the same scheduling set and the same reflection coefficients.
This property is difficult to learn by a conventional fully connected neural network, but is embedded in the architecture of a GNN\cite{PIandPE}. The
earlier work \cite{taojournal} shows that such a GNN can be trained to generate
interpretable RIS configurations and BS beamformers, but only for the setting
in which the user schedule is fixed a priori. In this paper, we treat the more challenging setting in which the user schedule also needs to be optimized. 

\begin{figure*}
\centering
\captionsetup{justification=centering}
\begin{subfigure}{0.75\textwidth}
\includegraphics[width=\textwidth]{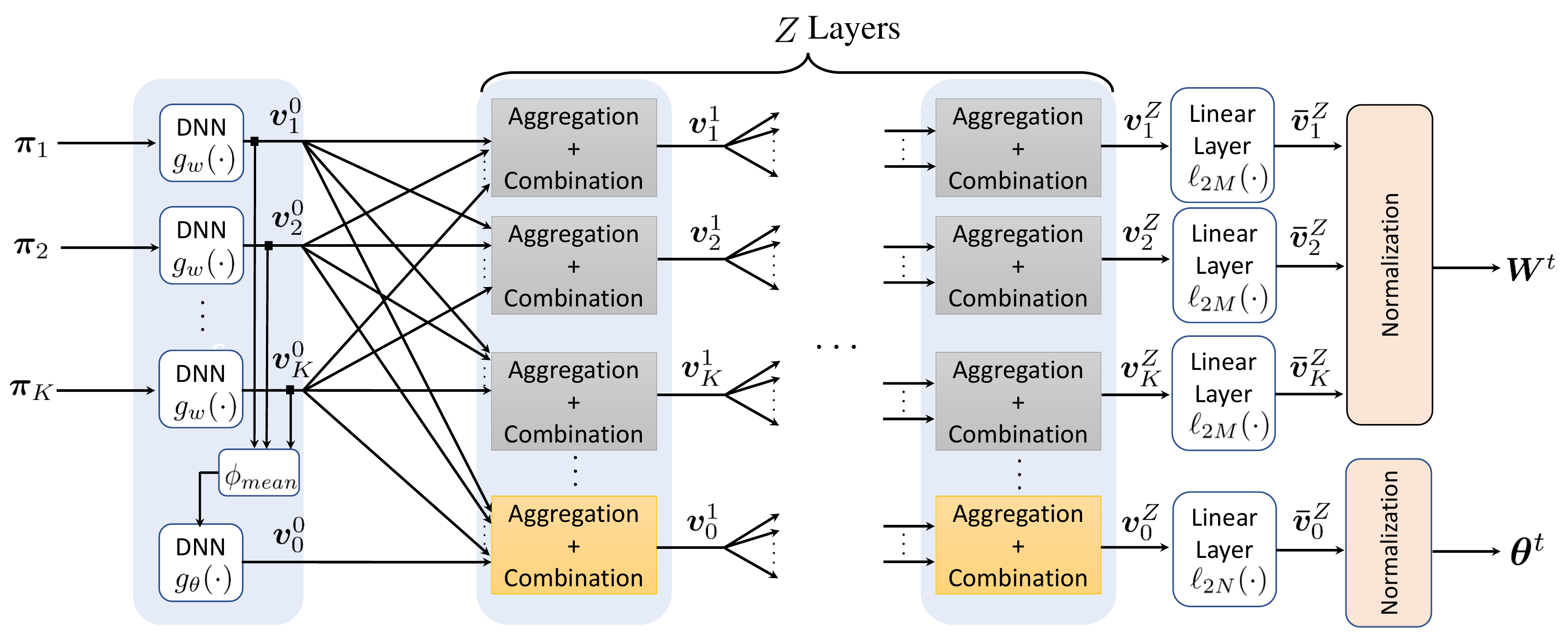}%
\caption{GNN Architecture used in Scheduling Stage.}%
\label{fig.gnnarchitecture}%
\end{subfigure}
\begin{subfigure}{0.57\columnwidth}
\includegraphics[width=\columnwidth]{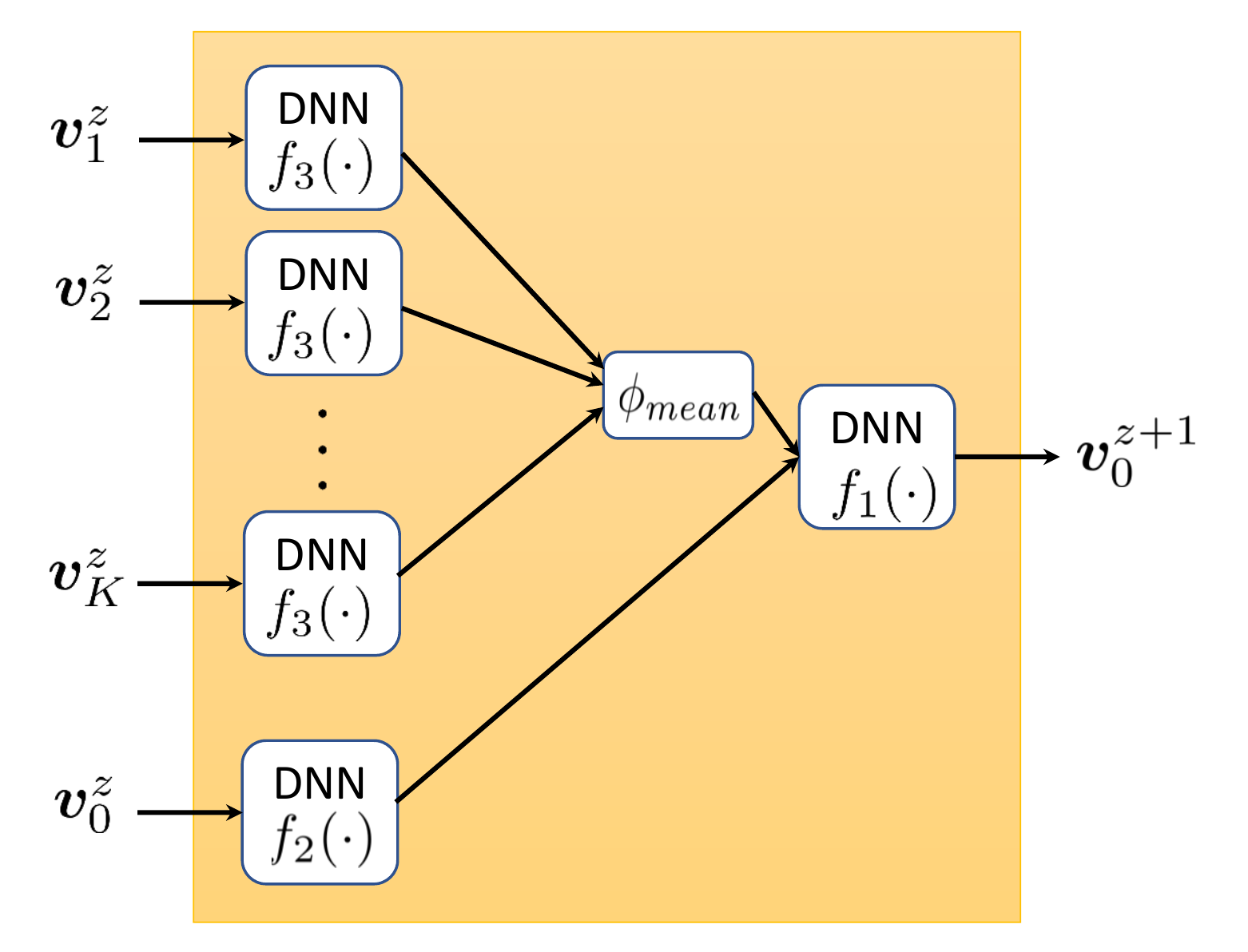}%
\caption{Aggregation and Combination Operation at the RIS node.}%
\label{fig.risnode}%
\end{subfigure}
\begin{subfigure}{0.57\columnwidth}
\includegraphics[width=\columnwidth]{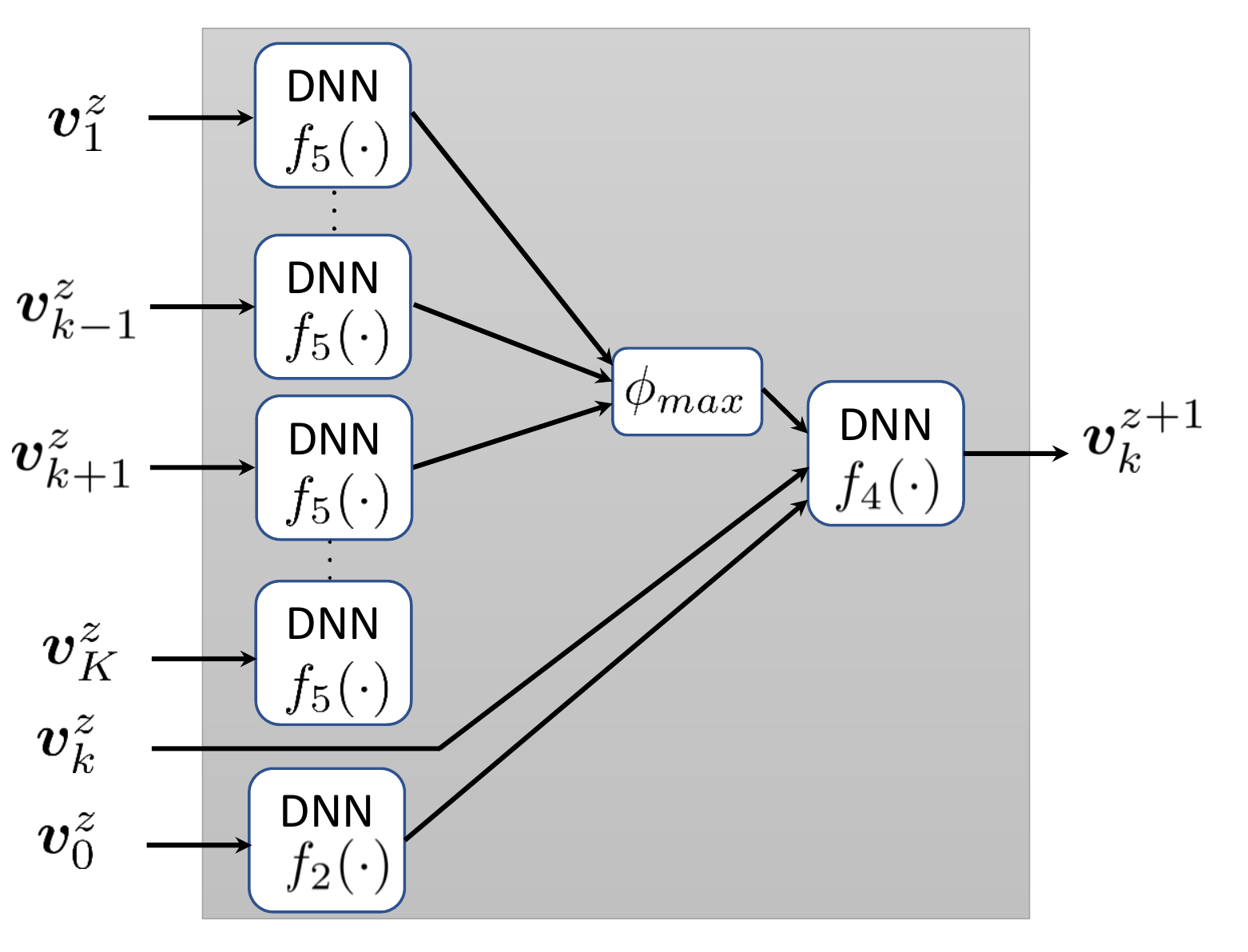}%
\caption{Aggregation and Combination Operation at the user node.}%
\label{fig.usernode}%
\end{subfigure}
\caption{GNN Architecture for $K$ users.}
\label{label:gnnarchitecture}
\end{figure*}

Toward this end, we adopt a GNN architecture, as shown in Fig.~\ref{label:gnnarchitecture}, which consists of $K+1$ fully connected nodes. Node $0$ corresponds to the RIS, and node $1$ to $K$ corresponds to beamforming vectors of user $1$ to user $K$. Each node is associated with a representation vector denoted as $\bm{v}_k, k\in \{0,...,K\}$. The idea is to encode useful information about each node in the representation vectors. The initial representation vectors $\bm{v}_k^0$ are functions of the input features $\bm{\pi}_k$'s, which include the weight and the received pilots of the user $k$ over $D$ sub-frames\begin{subequations} \label{vk_update}
\begin{align}
\bm{v}_0^0 = \;& g_\theta(\phi_{\rm mean} (\bm{\pi}_1,\dots,\bm{\pi}_K)),\\
\bm{v}_k^0 = \;& g_w(\bm{\pi}_k), k = 1,\cdots, K,
\end{align}
\end{subequations}
where $\bm{\pi}_k = [\alpha_k;\text{vec}(\mathcal{R}(\bm{\tilde{Y}}_k^{D}));\text{vec}(\mathcal{I}(\bm{\tilde{Y}}_k^{D}))]$ and $\phi_{\rm mean}$ is the element-wise mean function. Here, $g_w(\cdot)$ and $g_{\theta}(\cdot)$ are fully connected neural networks.

Then the nodal representation vectors are updated layer by layer, taking the representation vectors in the previous layer as input. Thus, the update of the RIS node is a function of itself and all user nodes; similarly, the update of a user node is a function of itself, the RIS node and all other user nodes. 
This updating rule allows the GNN to learn the interference amongst users. Specifically, the update rule in the $(z+1)$-th layer is given as \cite{taojournal}
\begin{subequations} 
\begin{align}
\bm{v}_0^{z+1} = \;& f_1\left(f_2(\bm{v}_0^z),\phi_{\rm mean}(f_3(\bm{v}_1^z),\cdots,f_3(\bm{v}_K^z))\right),\\
\bm{v}_k^{z+1} = \;& f_4\left(\bm{v}_k^z,f_2(\bm{v}_0^z),\phi_{\rm max}(\{f_5(\bm{v}_j^z)\}_{\forall j\neq 0,j\neq k})\right),
\end{align}
\end{subequations}
where $f_1(\cdot),f_2(\cdot),f_3(\cdot),f_4(\cdot),f_5(\cdot)$ are fully connected neural networks. Here, $\phi_{\rm max}$ is the element-wise max functions.

After $Z$ iterations, the final representation vectors $\bm{v}_k^Z$'s would contain the right representation of information to design the RIS configuration and the beamformers. The $\bm{v}_k^Z$'s pass through linear layer with $2N$ or $2M$ fully connected units
\begin{subequations}
\begin{align}
\bm{\bar{v}}_0^{Z} = \;& \ell_{2N}(\bm{v}_0^{Z})\in \mathbb{R}^{2N\times 1},\\
\bm{\bar{v}}_k^{Z} = \;& \ell_{2M}(\bm{v}_k^{Z})\in \mathbb{R}^{2M\times 1},~k = 1,\cdots,K.
\end{align}
\end{subequations}
Subsequently, $\bm{\bar{v}}_k^{Z}$'s are normalized so that the RIS reflection coefficients can be deduced from $\bm{\bar v}_0^Z$ and the beamforming vector associated with the user $k$ can be deduced from $\bm{\bar v}_k^Z, k\in \{1,\cdots,K\}$ as in \cite{taojournal}. 

The specific GNN architecture adopted here is similar to the one in
\cite{taojournal} in which the constituent components all obey permutation
invariant and equivariant properties, but with a key difference that the
proportional fairness weights for all the users are also used as input to the
GNN. Incorporating priority weights in the neural network for scheduling is in
general highly non-trivial, because the optimized system parameters can be very
sensitive to small perturbations in the weights \cite{Cui_2019_spatialdeep}. 
To tackle this issue, this paper proposes to use two GNNs, one for scheduling
and one for RIS configuration in order to optimize the overall objective.

%For this reason, this paper adopts a three-stage framework to first derive the user schedule using one GNN, then the RIS configuration using another GNN, and finally the beamformers, in separate stages.

%This property is difficult to learn by a conventional fully connected neural network, but is embedded in the architecture of a GNN. For example, GNN has shown excellent performance in network utility maximization 
%problems\cite{taojournal,gnn_sumrate1,gnn_sumrate2} by exploiting the finding that the information-sharing scheme of GNN determines the permutation invariant and equivariant property\cite{PIandPE}.

%Due to the universal approximation of deep neural network \cite{functionapprox}, we leverage deep neural network as a powerful function approximator to $f(\cdot)$ in problem (\ref{bigobjective}), i.e.,

More specifically, training a single GNN both to perform scheduling and to find 
the optimal RIS configuration is challenging, because such a GNN would need 
to have $K+1$ nodes, where $K$ can be large in a dense network, and each node 
would be associated with high-dimensional input features over the $D_\theta$ 
pilot sub-frames. 
%To address this scalability issue, we propose the following three-stage framework.
Instead, we propose to use a GNN with $(K+1)$ nodes over $D_\beta$ pilot 
sub-frames in the first stage just to produce the scheduling of up to $M$ users, 
where $D_\beta$ can be as small as $1$ to make the training manageable. 
Then, another GNN with $(M+1)$ nodes over $D_\theta$ pilot sub-frames is
adopted in the second stage to produce the optimized RIS configuration, where
$D_\theta$ can be considerably larger than $D_\beta$, but the pilots are
re-used so that the overall pilot length is $D_\theta$. We describe the two
GNNs in more detail below.

\begin{figure*}[!t]
  \includegraphics[width=\textwidth]{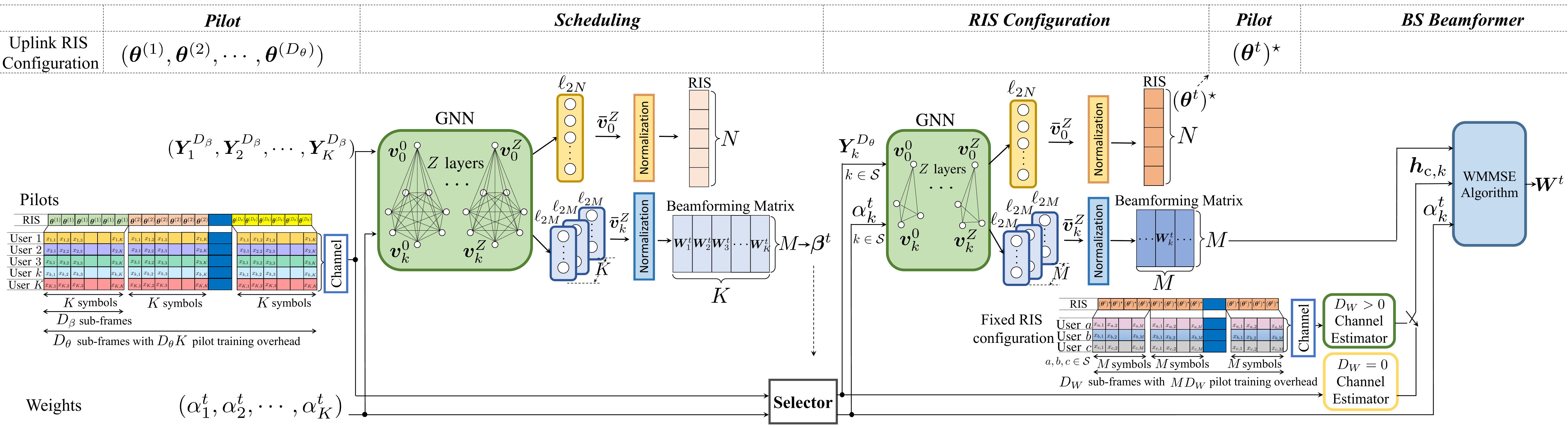}%
  \caption{The overall three-stage framework for scheduling, RIS configuration, and beamforming in multiuser downlink.}
  \label{fig.framework}%
\end{figure*}
\subsubsection{Scheduling}
In the first stage, a GNN with $K+1$ nodes is used to learn the optimized
schedule from a set of very short pilots of $D_\beta$ sub-frames. As shown in Fig.~\ref{fig.framework}, the inputs to the neural network are the user weights $(\alpha_1^t,\alpha_2^t,\cdots,\alpha_K^t)$ and the decorrelated received pilot sub-frames $(\bm{{Y}}_1^{D_\beta},\bm{{Y}}_2^{D_\beta},\cdots,\bm{{Y}}_K^{D_\beta})$. 
The output is the RIS configuration and beamformers designed for all $K$ users. 
The idea is that unlike BS beamforming or RIS configuration, which are strong functions of the
channel, scheduling can be done with only a coarse estimation of the overall
channel strength or directions. In practical implementation, the $D_\beta$
sub-frames can be part of the overall $D_\theta$ sub-frames of the received pilots,
since the channels are constant within the coherence time. 
%Specifically, the BS receives $D_{\theta}$ pilot sub-frames from all $K$ users  but only use $D_\beta$ out of $D_{\theta}$ received pilot sub-frames, i.e., $\tilde{\bm{Y}}_k^{D_\beta}$ are used as the input to the large GNN with $K+1$ nodes, as shown in Fig.~\ref{fig.framework}. 
In the training phase, the loss function is set to be the weighted sum rate of
$K$ users
\begin{equation}
    \begin{split}
    %\underset{\tiny \begin{array}{ll}(\bm W^t, \bm \theta^t) = \\ f(\{\bm Y^{(d)}\}_{d=1}^{D_{\beta}}, \bm \alpha^t)\end{array} }{\textrm{maximize}} 
    -\mathbb{E}\left[\sum_{k=1}^{K} \alpha_k^t R_k^t \right].
    \end{split}
\end{equation}

Since at this stage, we only use very short received pilots to train the GNN,  the reflection coefficients and the beamforming vectors learned by the GNN are quite suboptimal, but it is good enough to produce the scheduling decision. To do so, we adopt the implicit scheduling scheme and infer the schedule from the powers of the beamforming vectors $\bm w_k^t$'s learned by the GNN. Specifically, a user with larger beamforming power is more likely to be scheduled than those with smaller beamforming power. This is reasonable, because the GNN accounts for multiuser interference and designs the beamforming vectors together with the RIS reflection coefficients to maximize the objective \cite{taojournal}. Thus, the users to whom the GNN allocates more power are the users who make the most significant contributions to the weighted sum rate objective, and they should be included in the schedule. 
Therefore, if we select $M$ users with the highest powers from the set $\{\norm{\bm{w}^t_1}^2,\norm{\bm{w}^t_2}^2,\cdots,\norm{\bm{w}^t_K}^2\}$, this should be a good set of users to schedule. 

Note that the output $\bm{\theta}^t$ is discarded as in this stage our goal is to learn a scheduling decision using only $D_{\beta}$ received pilot sub-frames. 
Regardless, we still keep $\bm{\theta}^t$ in the training process to account for the contribution of RIS in the maximization of the objective. 
The reflection coefficients are designed in the next stage using another GNN that can map high-dimensional received pilots to a better $\bm{\theta}^t$.

\subsubsection{RIS Configuration}\label{gnnris}

In the second stage, we design better reflection coefficients using a second
GNN over the $M$ scheduled users but with longer $D_\theta$ pilot sub-frames. 
%The GNN is only with $M+1$ nodes corresponding to the $M$ scheduled users and the RIS, thus can
Such a GNN has a reduced number of nodes, i.e., $M+1$, so it can
take much longer received pilot sequences as input without experiencing
training difficulty. In particular, the inputs to the GNN for RIS
design are the user weights and $D_\theta$ decorrelated received pilot
sub-frames of the scheduled users, i.e., $a_k^t$ and ${\bm{Y}}_k^{D_\theta}$ with
$k\in\mathcal{S}$, as shown in Fig.~\ref{fig.framework}. The output is the RIS configuration and beamformers designed for the scheduled users. 
The GNN is trained to
maximize the weighted sum rate of the $M$ scheduled users to produce the optimized RIS configuration $(\bm\theta^t)^\star$. The loss function is expressed as
\begin{equation}
    - \mathbb{E}\left[\sum_{k\in\mathcal{S}} \alpha_k^t R_k^t \right].
\end{equation}
The designed RIS reflection coefficients $(\bm{\theta}^t)^*$ are employed as the RIS configuration for the scheduled users, but the beamformers are discarded and re-optimized in the next stage. 

The overall three-stage framework is shown in Fig.~\ref{fig.framework}.
Note that in the first two stages, the GNN for scheduling has 
larger dimension than the GNN for RIS configuration. But the GNN for RIS 
configuration takes considerably longer pilots than the GNN for scheduling.
\vspace{-1em}
\subsection{Beamforming Stage}\label{subsec:three-stage-framework}
Although the BS beamforming vectors are already produced in the
previous stages, there is still large room to fine-tune the beamforming
vectors. %The reason is that of the high dimensionality of the function mapping due to user weights\cite{Cui_2019_spatialdeep}, which imposes learning challenges for the neural network.

For example in the high SNR regime, an optimal beamforming design should closely resemble a zero-forcing (ZF) solution that can eliminate all the interference. %, rather than a simple matched filtering (MF) solution that aims to improve the channel gain. 
But a neural network may experience difficulties in learning a ZF beamformer in the high SNR regime as compared to learning a matched-filtering solution as shown in \cite{zfmf}. 
%Aside from neural network, with the RIS configuration fixed, the effective combined channel between the BS and the scheduled users is of much lower dimension, so estimating the combined channel between the BS and scheduled users requires a much smaller pilot training overhead.

For this reason, we propose to fine-tune the beamforming matrix using the WMMSE algorithm based on the estimated channel $\bm{h}_{{\rm c},k}$ in a third stage. Specifically, we use the following iterative updating rule to obtain an stationary solution \cite{wmmse}:
\begin{subequations}
\begin{align}
\xi_k = \;& \left(\sum_{j\in \mathcal{S}} |\bm{ h}_{{\rm c},k}^{\sf H}\bm{w}_j^t|^2+\sigma_d^2\right)^{-1}\bm{ h}_{{\rm c},k}^{\sf H}\bm{w}_k^t,\\
\nu_k = \;& \left(1-\xi_k^*\bm{ h}_{{\rm c},k}^{\sf H}\bm{w}_k^t\right)^{-1},\\
\bm{w}_k^t = \;& \alpha_k^t\xi_k\nu_k \left(\lambda \bm{\text{I}}_M+\sum_{j\in \mathcal{S}}\alpha_j^t|\xi_j|^2\nu_j\bm{ h}_{{\rm c},j}\bm{ h}_{{\rm c},j}^{\sf H}\right)^{-1}\bm{ h}_{{\rm c},k},
\end{align}
\end{subequations}
where $\lambda$ denotes the dual variable for the downlink transmission power constraint.

To estimate the channel of the $M$ scheduled users, we can use one of the two channel estimation schemes discussed in Section~\ref{sec:pilot_beamformer}.
We could either employ additional pilots to estimate the effective low-dimensional channel, or estimate the high-dimensional channel without additional pilot overhead. 

\subsection{Pilot Overhead}

The overall pilot overhead of the three-stage framework is as follows. 
The scheduling and RIS configuration stages take $D_\theta$ pilot sub-frames, each
of length $K$. Then, we have two cases depending on the two strategies in the
beamforming stage.

\subsubsection{With Additional Pilots ($D_W > 0$)} \label{withadditionalpilot}

In this case, the scheduled users transmit additional orthogonal pilots of length $M$ over the channel with fixed uplink RIS configuration $(\bm\theta^t)^\star$. 
%Note that perfectly estimating the combined channel vectors $\bm h_{{\rm c},k} $ of $M$ scheduled users requires only $M$ pilots in the noiseless scenario. 
Depending on the SNR, we can use $M \times D_W$ pilot sequences. This CSI acquisition strategy allows us to obtain an accurate estimation of the low-dimensional channel with a small pilot overhead for each scheduling timeslot. %Let $\Upsilon$ denote the total number of scheduling timeslots in the channel coherence period. 
In this case, the overall pilot overhead over all three stages is given by
\begin{equation}\label{LTcalculation}
L=K \times D_{\theta} + M \times D_W \times \Upsilon,
\end{equation}
where $K \times D_{\theta}$ accounts for the pilots used to design the schedule and the RIS configuration in the first two stages, and $\Upsilon$ is the number of scheduling slots.

\subsubsection{Without Additional Pilots ($D_W = 0$)} 
Alternatively, we can re-use the received $D_{\theta}$ pilot sub-frames from all $K$ users to estimate the direct channel $\bm{h}_{{\rm{d}}, k}$ and cascade channel $\bm{A}_k$ for the scheduled users. In this case, the total pilot overhead is
\begin{equation} 
L=K \times D_{\theta}.
\end{equation}
In the next section, we provide a comparison of the two cases in terms of performance versus the overall pilot overhead in various scenarios.

\section{Numerical results}
\label{sec.numerical_irs}
\subsection{Simulation Environment}

\begin{figure}[!t]
  \includegraphics[width=\columnwidth]{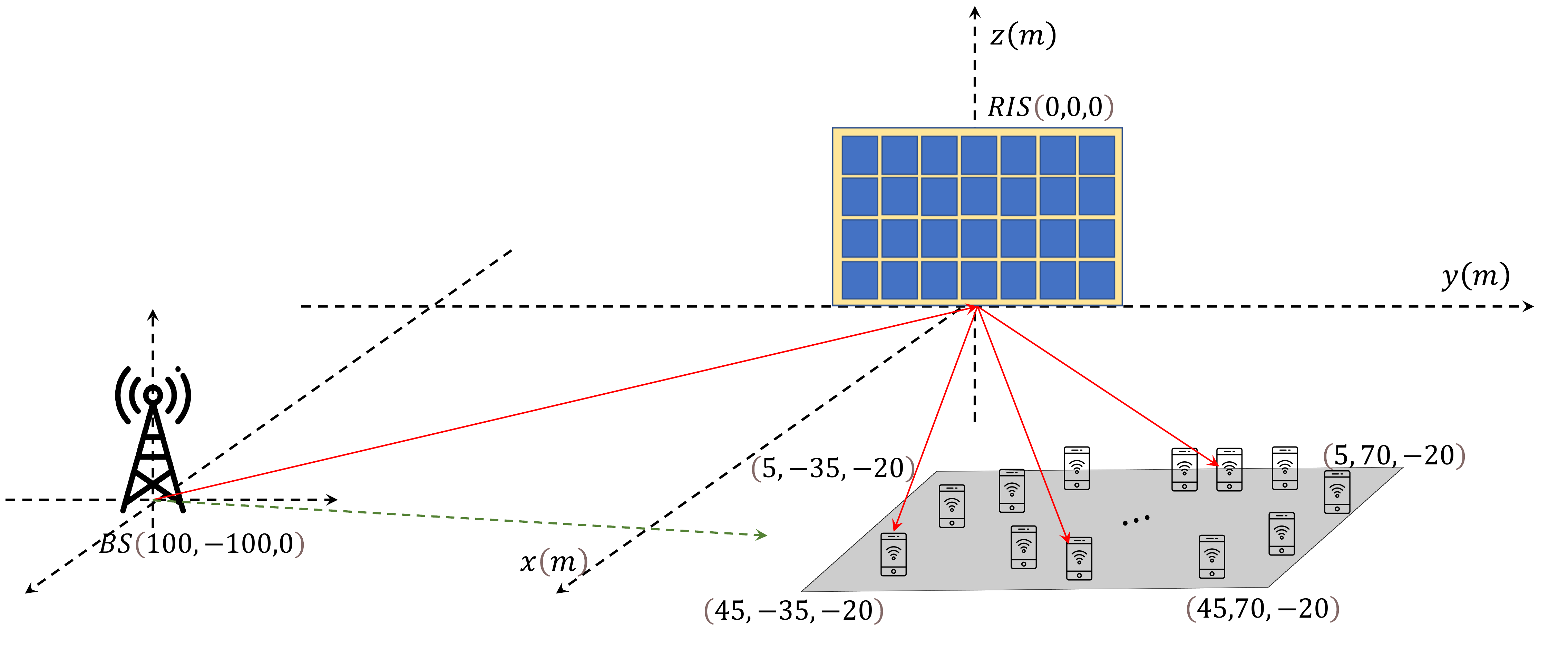}%
  \caption{Simulation setting for RIS-assisted multiuser network.}
  \label{fig.simulationsetting}%
\end{figure}

We consider an RIS-assisted multiuser MISO network with $M=8$ BS antennas, $N=128$ RIS reflective elements, and $K = 32$ users. 
In the $(x,y,z)$ coordinates, the BS and the RIS are located at $(100m,-100m,0m)$ and $(0m, 0m, 0m)$ respectively. The user locations are uniformly generated within a rectangular area on the $x$-$y$ plane $(25\pm20m, 17.5\pm52.5m, -20m)$, as shown in Fig.~\ref{fig.simulationsetting}. We assume that the direct link channel follows Rayleigh fading
\begin{equation} 
\bm{h}_{\rm d,k} = \rho_{0,k}\bm{\tilde{h}}_{\rm d,k},
\end{equation}
where $\bm{\tilde h}_{{\rm d},k}\sim \mathcal{C}\mathcal{N}(0,\bm{I})$ and $\rho_{0,k}$ denotes the pathloss between BS and user $k$. The reflection channels $\bm{h}_{{\rm r},k},\bm{G}$ are assumed to follow Rician fading model:
\begin{subequations} 
\begin{align}
\bm{h}_{{\rm r},k} &= \rho_{1,k}\left(\sqrt{\dfrac{\epsilon}{1+\epsilon}} \bm{\tilde{h}}_{{\rm r},k}^{\rm LOS} + \sqrt{\dfrac{1}{1+\epsilon}}\bm{\tilde{h}}_{{\rm r},k}^{\rm NLOS}\right),\\
\bm{G} &= \rho_2\left(\sqrt{\dfrac{\epsilon}{1+\epsilon}}\bm{\tilde{G}}^{\rm LOS} + \sqrt{\dfrac{1}{1+\epsilon}}\bm{\tilde{G}}^{\rm NLOS}\right),
\end{align}
\end{subequations}
where $\rho_{1,k}$ and $\rho_2$ denote the path losses between the RIS and the $k$-th user/BS. The path-loss models of the direct and reflected paths are $32.6+36.7\log(d_1)$ and $30+22\log(d_2)$, respectively, where $d_1$ and $d_2$ denote the corresponding link distance.
Here,
$\bm{\tilde{h}^{\textrm{NLOS}}}_{{\rm r},k}$ and $\bm{\tilde{G}^{\textrm{NLOS}}}$ denote the non-line-of-sight components and their entries are generated independently according to $\mathcal{C}\mathcal{N}(0,1)$. We assume that there are $\Upsilon=50$ scheduling timeslots in slow-fading channel and $\Upsilon = 5$ scheduling timeslots in fast-fading channel. The transmission power for uplink and downlink are $15$dBm. The bandwidth is $10$MHz with a background noise of $-170$dBm/Hz. The Rician factor $\epsilon$ is set to $10$. The forgetting factor $\gamma$ in exponentially moving averaging of user rates is set to $0.01$.
%We implement two-layer ($Z=2$) GNN models using Tensorflow \cite{tensorflow}; the GNN is trained by Adam optimizer \cite{adam}. 

\subsection{Baseline Scheduling Strategy}
As baseline comparisons, the following greedy scheduling (GS) algorithm and exhaustive search (ES) scheduling algorithm 
for solving problem (\ref{bigobjective}) are implemented. 

The greedy scheduling algorithm consists of two phases: an uplink channel estimation phase and a downlink proportional fairness scheduling phase.
%We assume channel reciprocity in a TDD system such that the channel estimate of the uplink direction can be directly used for downlink transmission. Thus, 
Uplink channel estimation is performed using the method in \cite{jiechannelest}, and the downlink weighted sum-rate maximization problem is solved with a greedy scheduler and a block coordinate descent (BCD) approach for optimizing the beamformers and the RIS configuration as in \cite{hguo}.

\subsubsection{Uplink Channel Estimation}\label{lmmse_irs}

Channel estimation is performed at the beginning of channel coherence period, in which the BS collects all of the pilot symbols from $K$ users. Thus, the total pilot overhead for a coherence interval is calculated as 
\begin{equation} 
L=K \times D_H,
\end{equation}
where $D_H$ denotes the number of pilot sub-frames used to estimate the high-dimensional channels $\bm{A}_k$'s and $\bm{h}_{{\rm d},k}$'s via 
the LMMSE estimator \eqref{lmmse}. 
%$\bm{{H}}_k$'s. %Subsequently, the estimated direct channel $\bm{{h}}_{{\rm d},k}$ and cascade channel $\bm{{A}}_k$ can be recovered from $\bm{{H}}_k$. 

%For the conventional two-step approach, channel estimation which maps received $D_H$ pilot frames to CSI is performed at the beginning of each channel coherence time block. Subsequently, based on estimated CSI, scheduling and beamforming is performed for each scheduling timeslot. Let $L_T$ denotes the total pilot overhead in symbols for a coherence time block. In the case of the baseline approach:

%Though the linear restriction imposed on the estimator $h$ may result in suboptimal solution to (\ref{mmse}), it is optimal if the underlying $\bm{H}_k$ is Gaussian distributed.

\subsubsection{Downlink Proportional Fairness Scheduling}\label{baselinealgo}

The joint scheduling, RIS configuration, and beamforming problem can be solved iteratively as described in Algorithm \ref{ris_bl}. The user scheduling is based on greedily choosing the user that maximizes the overall optimization objective of weighted sum rate. 

In the simulations, we evaluate the performance of the proposed data-driven approach against the baseline approach for the cases both with perfect CSI and with estimated CSI. 

As another baseline comparison, the ES scheduling algorithm tries every possible combination of users to identify the optimal scheduling. For each combination of users, the RIS configuration is designed using the GNN in Section \ref{gnnris}, and beamforming matrix is designed using WMMSE algorithm as in Section \ref{subsec:three-stage-framework}. The set of users that achieves the highest in objective (\ref{bigobjective}) is the scheduled user set. 
The ES scheduling algorithm is not scalable as the number of users becomes large. It serves as a performance upper bound for the proposed GNN-based scheduler. 
\begin{algorithm}[t]
  \caption{Baseline GS with BCD}
    \label{ris_bl}
  \begin{algorithmic}[1]
    \item Initialize the set of scheduled users $\mathcal{S}=\{\emptyset\}$;
    \item Initialize random RIS phase shift vector $\bm{\theta}^t$;
    \item Set $\bm{w}_k^t = (\bm{{h}}_{{\rm{d}},k}+ \bm{{A}}_k\bm{\theta}^t), k = 1,\cdots,K$; 
    \item Compute and sort weighted single-user rate $\alpha_k^t\hat{R}_k^t$ \linebreak where 
\begin{equation*} \label{singleuserrate}
\begin{aligned}
\hat{R}_k^t = \log\left(1 +\dfrac{|(\bm{{h}}_{{\rm{d}},k}+\bm{{A}}_k\bm{\theta}^t)^{\sf H} \bm{w}_k^t|^2}{ \sigma_d^2}\right);
\end{aligned}
\end{equation*}
\item Select the user with the largest $\alpha_k^t\hat{R}_k^t$ to add to $\mathcal{S}$;
\item \label{traverse} For each unscheduled user, test whether adding that \linebreak user to $\mathcal{S}$ improves the objective (\ref{bigobjective}). Select the user with the largest improvement to add to $\mathcal{S}$. Repeat until \linebreak adding another user no longer improves the objective.
\item \label{bl_updateris} Fix $\mathcal{S}$ and $\bm{W}^t$, update $\bm{\theta}^t$ using the Riemannian conjugate gradient (RCG) algorithm \cite{RCG};
\item \label{bl_updatew} Fix $\mathcal{S}$ and $\bm{\theta}^t$, update $\bm{W}^t$ using WMMSE \cite{wmmse};
\item \label{bl_updatebeta} Fix $\bm{\theta}^t$ and $\bm{W}^t$, update $\mathcal{S}$ by adding the unscheduled user which improves the objective (\ref{bigobjective}) the most, if any; 
\item Repeat step \ref{bl_updateris}-\ref{bl_updatebeta} until convergence.
  \end{algorithmic}
\end{algorithm}
\begin{figure*}[hbtp]
\centering
\begin{subfigure}{\columnwidth}
\includegraphics[width=\columnwidth]{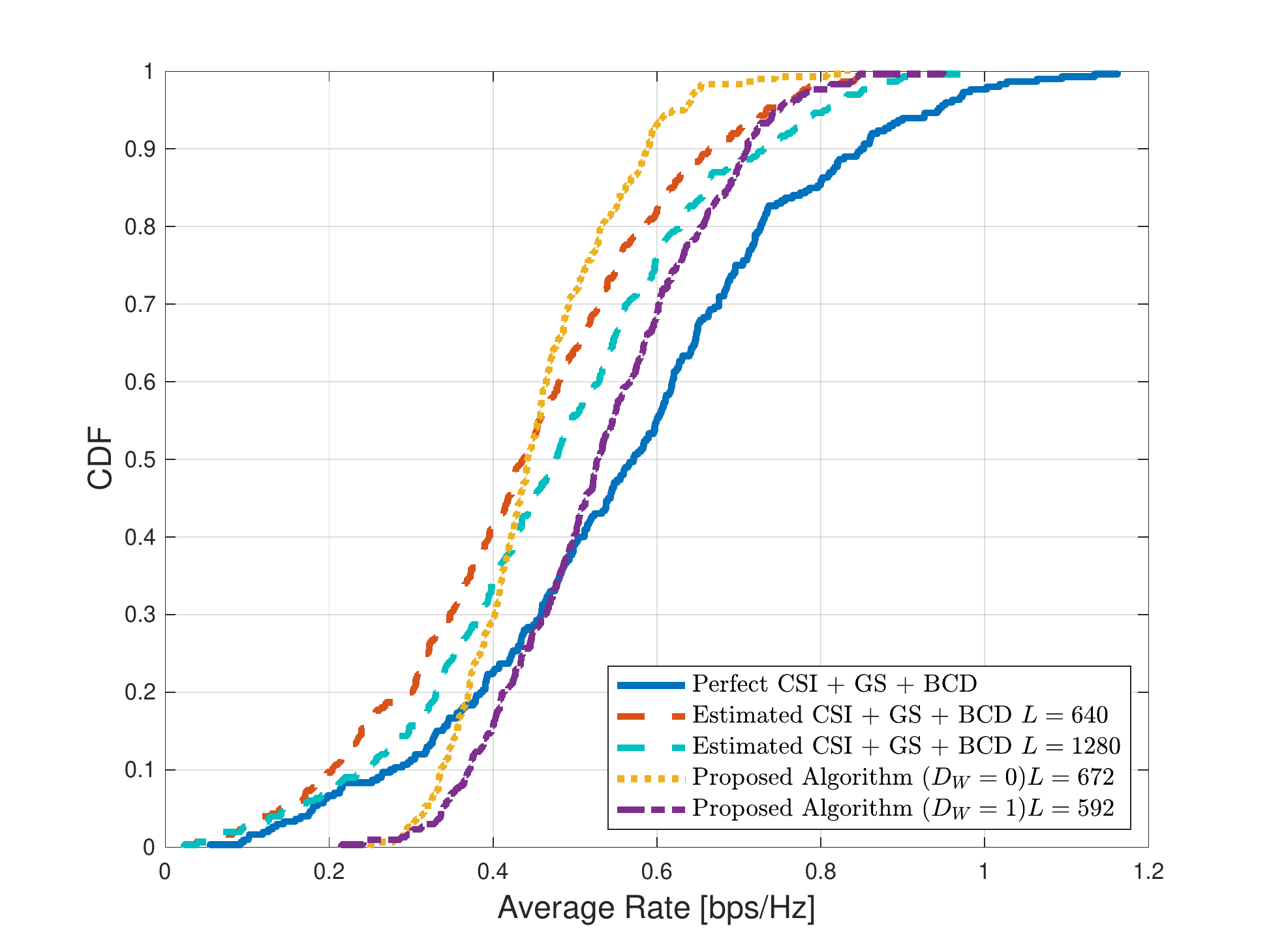}%
\caption{Cumulative distribution function (CDF) of user rates in slow-fading channel.}%
\label{fig.slowfading_cdf}%
\end{subfigure}
\begin{subfigure}{\columnwidth}
\includegraphics[width=\columnwidth]{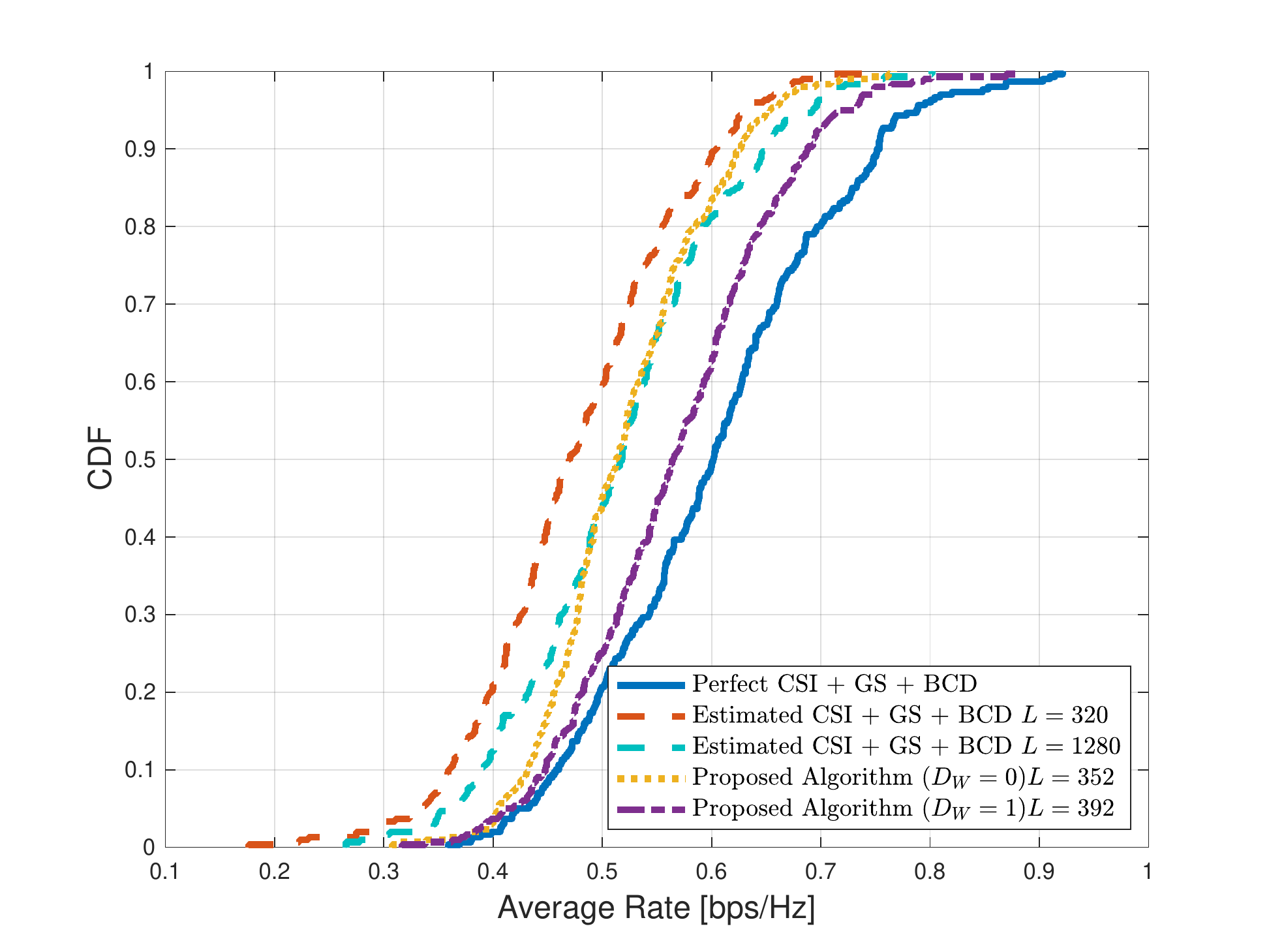}%
\caption{Cumulative distribution function (CDF) of user rates in fast-fading channel.}%
\label{fig.fastfading_cdf}%
\end{subfigure}
\begin{subfigure}{\columnwidth}
\includegraphics[width=\columnwidth]{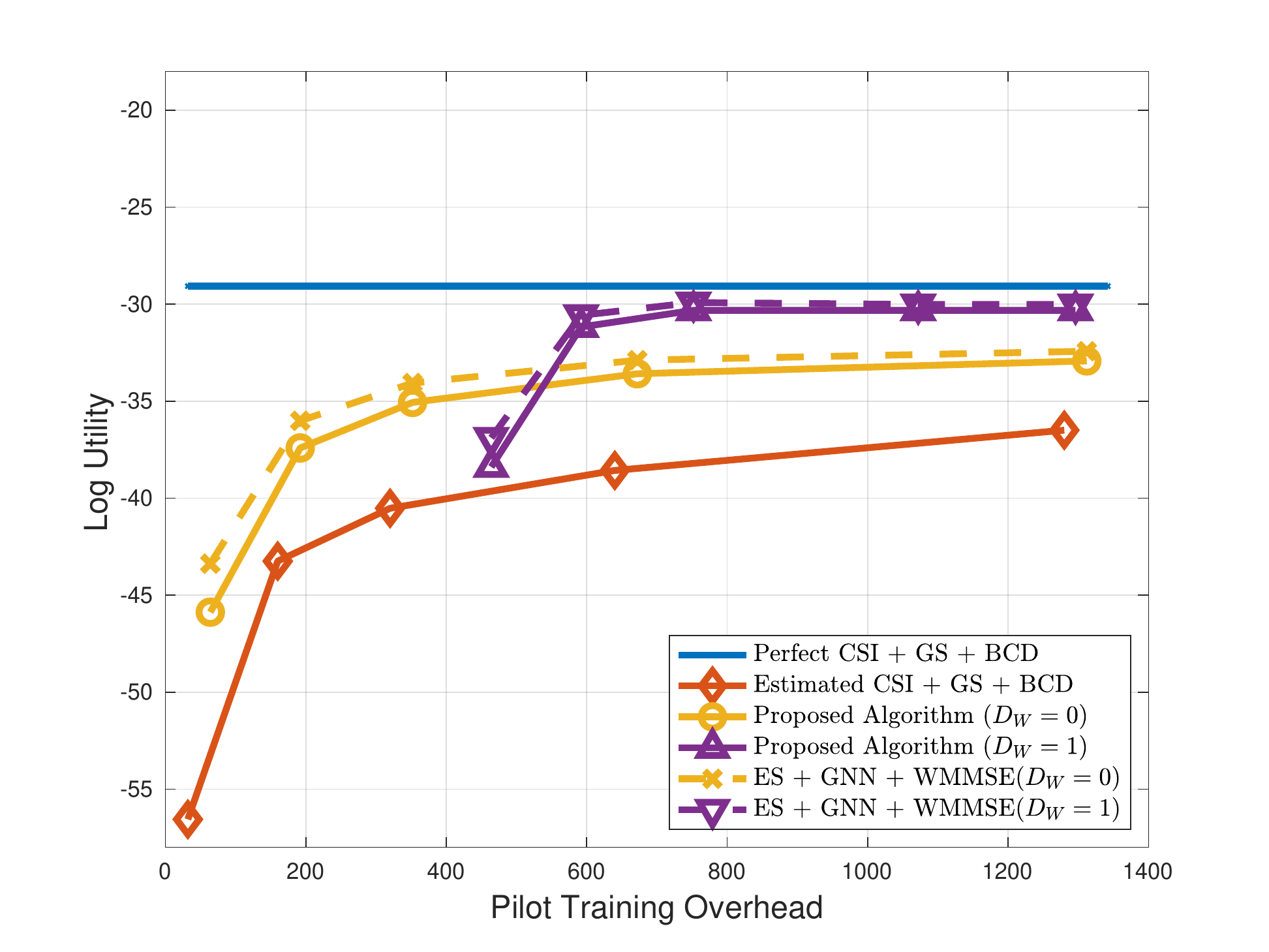}%
\caption{Network-wide log utility of slow-fading channel.}%
\label{fig.slowfading_logu}%
\end{subfigure}
\begin{subfigure}{\columnwidth}
\includegraphics[width=\columnwidth]{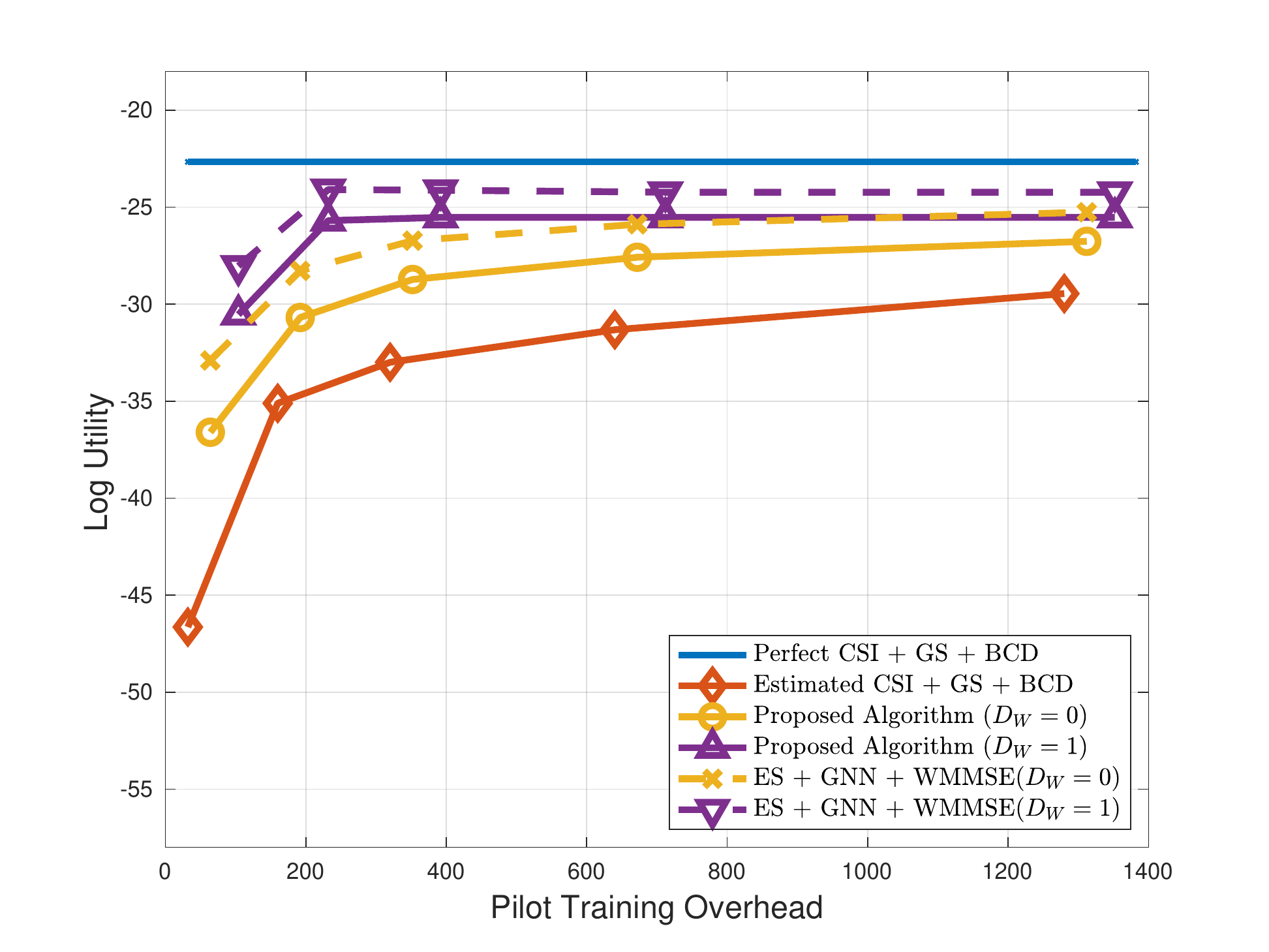}%
\caption{Network-wide log utility of fast-fading channel.}%
\label{fig.fastfading_logu}%
\end{subfigure}
\begin{subfigure}{\columnwidth}
\includegraphics[width=\columnwidth]{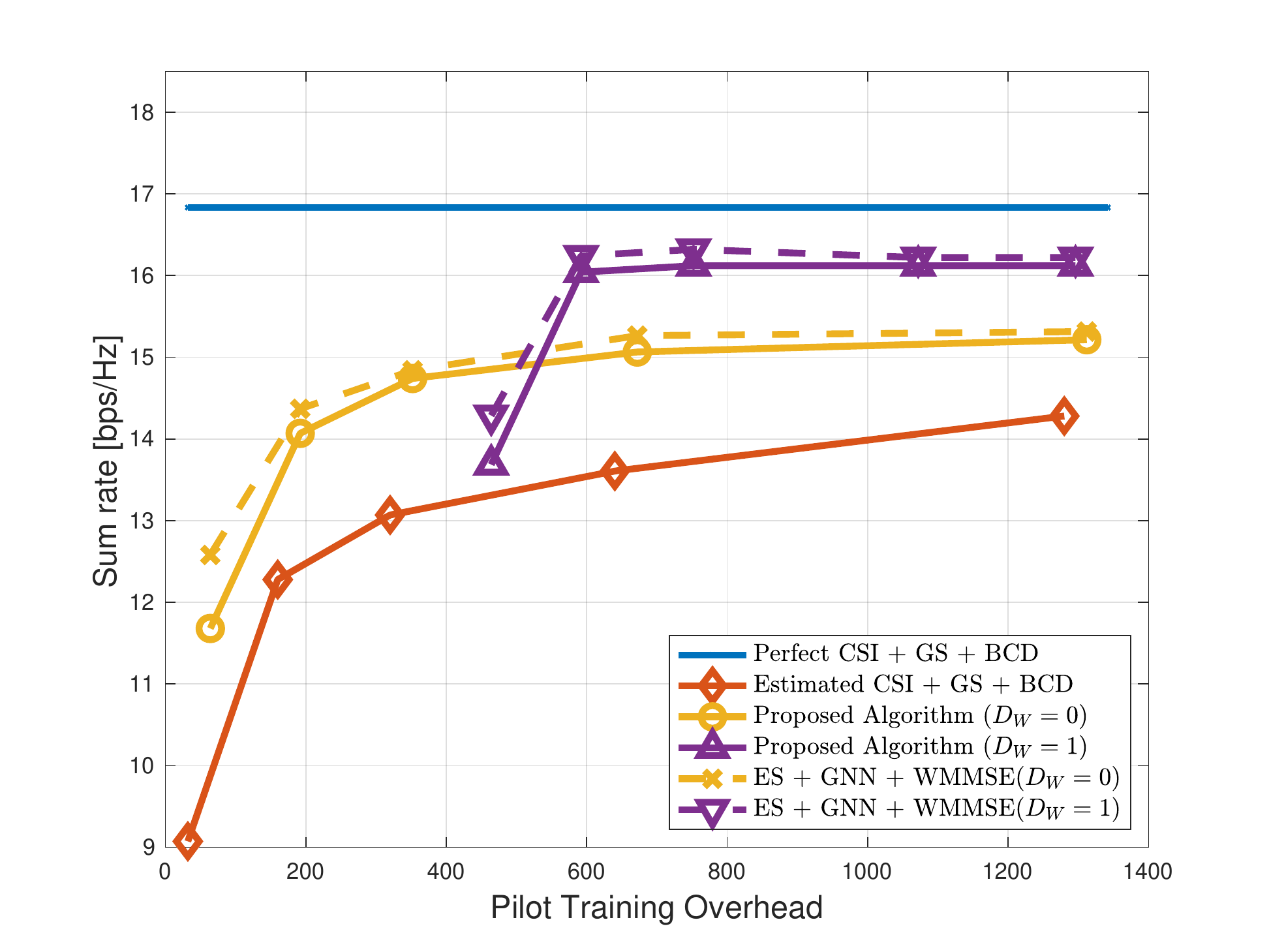}%
\caption{Sum rate of slow-fading channel.}%
\label{fig.slowfading_sumrate}%
\end{subfigure}
\begin{subfigure}{\columnwidth}
\includegraphics[width=\columnwidth]{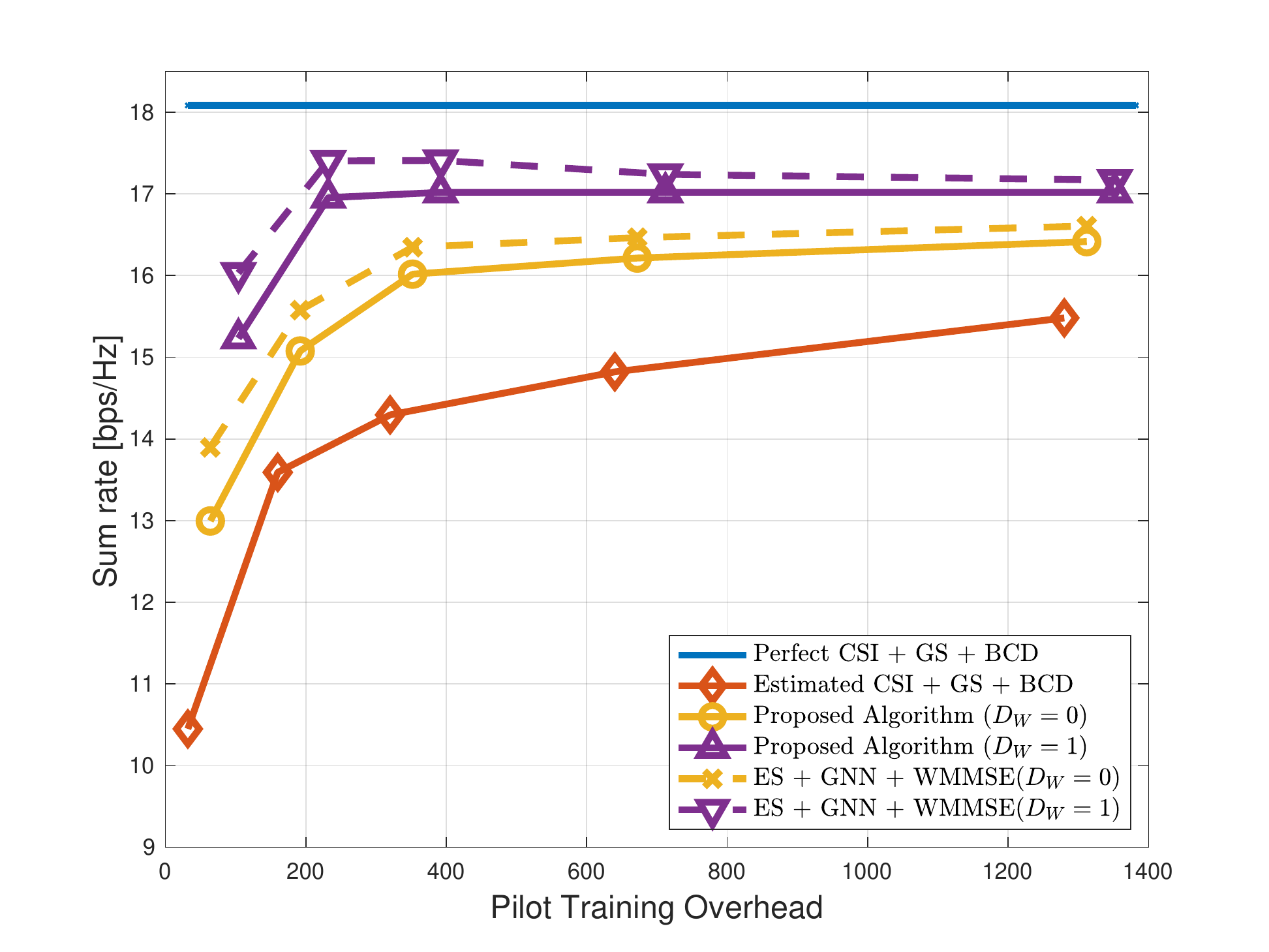}%
\caption{Sum rate of fast-fading channel.}%
\label{fig.fastfading_sumrate}%
\end{subfigure}
\caption{Performance of GNN based scheduling in an RIS-assisted downlink system with $N=128$, $M=8$, $K=32$, $P_d=15$dBm.}
\label{cdflogsum}
\end{figure*}

\subsection{Simulation Result}

The GNN based three-stage joint scheduling, RIS configuration, and beamforming framework as shown in Fig.~\ref{fig.framework} is
implemented using parameters in Table \ref{Tab:fcnn}. 
For both the GNN for scheduling and the GNN for RIS configuration, we implement a two-layer ($Z=2$) models using Tensorflow \cite{tensorflow}. %, and the GNN is trained using Adam optimizer \cite{adam}. 
In the scheduling stage, we set $D_{\beta} = 1$. 
In the training phase, the GNNs sample 102,000 training data including the channel vectors, pilot sequences, and uniformly generated weights in each epoch.
%is fed with training data including training channel vectors, training pilot sequences, and uniformly generated training weights. 
%$\bm{h}_{{\rm d},k}^{train}$, $\bm{A}_{k}^{train}$, $\bm{\tilde{Y}}_k^{train}$, and uniformly generated $a_k^{train}$, $\forall k$. 
The Adam optimizer \cite{adam} is used for the neural network to adjust its neuron weights in an unsupervised fashion to minimize the loss function. 
%The learning phase stops until the mapping from the received pilots and weights to target output is learnt. 
%In the testing phase, the trained neural network can then generate the desired output, i.e., ($\bm{\theta}^t,\bm{W}^t$), from received pilot sequences and weights.
We examine its performance in both slow-fading and fast-fading channel environments below. 
\begin{table}[t]
\begin{center}
\captionof{table}{Hyperparameters of Deep Neural Networks\label{Tab:fcnn}}
\begin{tabular}{ |c|c|c| } 
\hline
\textbf{\shortstack{Label\\\;}}& \textbf{\shortstack{Activation\\Function}}& \textbf{\shortstack{Size\\\;}}\\
\hline
$\ell_{2N}(\cdot)$  &  Linear & $2N$\\
\hline
$\ell_{2M}(\cdot)$  &  Linear & $2M$\\
\hline
$g_w(\cdot)$  &  Relu & $2MD_{\theta}\times1024\times 512$ \\
\hline
$g_\theta(\cdot)$  &  Relu & $512 \times 1024 \times 512$ \\
\hline
\shortstack{$f_1(\cdot),f_2(\cdot),$\\$f_3(\cdot),f_4(\cdot),f_5(\cdot)$} &  \shortstack{Relu\\ \quad\\ \quad} & \shortstack{$512 \times 512 \times 512$ \\\quad  }\\
\hline
\end{tabular}
\end{center}
\end{table}

\subsubsection{Slow-fading channel}

We first examine the performance of GNN based algorithms vs. the baseline in terms of the \textit{cumulative distribution function} (CDF) of user average rate in 
Fig.~\ref{cdflogsum}(\subref{fig.slowfading_cdf}). Each point on the CDF curve represents an individual user's average instantaneous rate averaged across $10^2$ i.i.d channel realizations, which is equivalent to $10^2\Upsilon$ scheduling timeslots. % with $10^3$ i.i.d channel realizations in total. 
%For each channel realization, we assume a channel coherence period with $\Upsilon = 50$ available scheduling timeslots in a slow-fading environment where up to $M = 8$ user can be served simultaneously in each time slot. In a fast-fading environment, the available number of scheduling timeslots reduces to $\Upsilon = 5$. We set $D_{\beta} = 1$ unless otherwise stated. 

From Fig.~\ref{cdflogsum}(\subref{fig.slowfading_cdf}), the proposed
data-driven algorithm with $592$ pilot symbols (including $D_W=1$ pilot 
per scheduling slot in the third beamforming stage) demonstrates performance close to the baseline approach with perfect CSI, and
significantly outperforms the baseline approach with estimated CSI using $640$ or $1280$ 
pilot symbols.
%Similarly, the proposed algorithm $(D_W = 0)$ with $672$ pilot symbols demonstrates comparable performance to the baseline approach with LMMSE estimated CSI over $1280$ symbols. 
This implies that the proposed algorithm,
which is based on directly maximizing system objective based on the received pilots and the user priority weights using a neural network without explicit channel reconstruction,
can significantly reduce the amount
of the pilot overhead as compared to the conventional approach. 

It is also interesting to see that the proposed algorithm
demonstrates a strong CDF performance in the low-rate regime (below
$40$-percentile range). This is because the scheduled users tend to have either
strong channels or high weights (if not both). A strong CDF performance in the
low-rate range implies that the GNN scheduler is sensitive to user weights and
often schedules users with high weights and weak channels.

%It is also interesting to see the proposed algorithm demonstrates a better CDF performance in the low-rate regime. This implies that the GNN scheduler is sensitive to user weights and attentive to users with poor quality channels.

Further, from Fig.~\ref{cdflogsum}(\subref{fig.slowfading_logu})
and Fig.~\ref{cdflogsum}(\subref{fig.slowfading_sumrate}),
we observe that the proposed algorithm but without additional pilot in the
beamforming stage also outperforms the baseline approach consistently, and in the
short pilot region, can be superior to the approach of using additional pilot
for beamforming in the third stage. The performance gain against the baseline
is again due to the more direct and efficient use of the received pilot sequences by the GNN models, without explicit channel estimation as the intermediary. 

Finally, we remark that a slight performance improvement can be obtained by the ES scheduling algorithm. However, exhaustive search is computationally complex and not scalable as the number of users becomes large.

We note that there is a performance gap between both proposed algorithms and
the perfect CSI case. For the case without the additional pilot in the beamforming
stage, the gap is due to the inaccurate estimation
of the high-dimensional channel $\bm{A}_k$, $\bm{h}_{{\rm d},k}$, $k = 1,\cdots,K$. The performance gap is reduced when additional pilots are
used to estimate the low-dimensional effective 
%once $D_W$ is set to be greater than 0, in which the fixed-RIS-configuration
%pilot transmission scheme enables an accurate estimation of the low-dimensional
channel $\bm{h}_{{\rm c},k}$, $k \in \mathcal{S}$. %thus a well-optimized beamformer. 
But the gap is still not zero. This is because of the inherent limitation
in the neural network architecture and training.

%Moreover, the same figures reveal that to accurately estimate all unknown channel coefficients in $\bm{G}$, $\bm{h}_{{\rm r},k}$, $\bm{h}_{{\rm d},k}$, at least 1000 pilot symbols are needed from all $K = 32$ users.  It is clear that the goal of accurate recovery of CSI inflates the need for pilots and hinders the convergence of the baseline algorithm. Conversely, the proposed algorithm $(D_W = 1)$ converges rapidly to the baseline approach with perfect CSI with $592$ pilot symbols from all $32$ users.  Based on those observations, we assert that directly maximizing system objective based on received pilots and weights can reduce pilots overhead drastically, as the need for accurate channel reconstruction is alleviated. 

\subsubsection{fast-fading channel}

In a fast-fading environment, the channel coherence period is shorter, thus 
there are fewer number of scheduling timeslots available. 
From Fig.~\ref{cdflogsum}(\subref{fig.fastfading_logu}) and
Fig.~\ref{cdflogsum}(\subref{fig.fastfading_sumrate}), we observe that 
using additional pilots in the beamforming stage is overall the best approach.
Note that both proposed approaches always outperform the baseline.

We also note that the log utility and the sum rate are higher in a fast-fading
setting as compared to a slow-fading setting. This is due to multiuser diversity.
In a slow-fading scenario, the channels remain fixed for a long period of time.
At first, the scheduler tends not to schedule users with weak channels. But after a
large number of scheduling timeslots, the scheduler has no choice but to
schedule those users, which can result in a lower sum rate.  This
phenomenon is not as prevalent in a fast-fading channel, where the user
channels change more rapidly and the scheduler is less likely to be forced to
schedule a user with a poor channel condition.

\subsection{Complexity Analysis}

The complexity of the proposed learning based approach consists of two neural network inference stages and an iterative beamformer optimization stage. 
The complexity of the GNN in the scheduling stage is ${O}(MK\tau +(K+1)^2\epsilon Z + (MK+N)\delta)$, %\cite{spectrumlearning,gnn}, 
where $\tau$, $\delta$, and $\epsilon$ denote the dimension of the input layer, dimension of the output layer, and the computational complexity of the fully connected aggregation and combination operations respectively. Recall that $Z$ denotes the number of layers of the GNN. 
Similarly, the complexity of GNN for the RIS design is ${O}(M^2\tau+(M+1)^2\epsilon Z + (M^2+N)\delta)$.
The complexity of the beamformer optimization stage is ${O}(\ell_{\lambda}\ell_w (2NM^2+M^3))$\cite{hguo}, where $\ell_{\lambda}$ and $\ell_w$ denote the iteration number of inner loops. The total complexity is therefore ${O}((K^2+M^2) Z + MK+M^2+N +\ell_{\lambda}\ell_w (NM^2+M^3))$ after discarding lower order terms. 
It is important to note that both the training and inference processes are highly parallelizable using modern graphic process units (GPUs), so that in practice the proposed approach can be executed very efficiently.
%The proposed learning based approach is computationally efficient in the simulations as the algorithm is one-pass. In addition, the GNN inference stage can be executed efficiently by GPU via parallel computing.

In contrast, the baseline GS strategy in Algorithm \ref{ris_bl} adopts an iterative structure that solves scheduling, RIS reflection coefficients and beamformer sub-problems sequentially until convergence. 
%The road to convergence is not instantaneous, in which during the process, significant computation resources are used to sequentially solve the sub-problems again and again. 
In particular, the complexities associated with calculating the schedule, the RIS reflection coefficients, and the beamforming vectors are respectively ${O}(MNK\log(K))$, ${O}(\ell_r M^2N^2)$, and ${O}(\ell_{\lambda}\ell_w (2NM^2+M^3))$\cite{hguo}, where $\ell_r$ denotes the iteration number of inner loops. The total complexity is ${O}( \ell_O(MNK\log(K)+\ell_r M^2N^2+\ell_{\lambda}\ell_w (NM^2+M^3)))$, where $\ell_O$ denotes the iteration number of the outer loop. In practice, both the GS and ES algorithms are orders of magnitude slower than the proposed neural network based approach.

%We remark that the baseline method, which adopts an iterative structure, is not as computationally efficient as the proposed one-pass learning based algorithm.  

\begin{figure}
\centering
\includegraphics[width=\columnwidth]{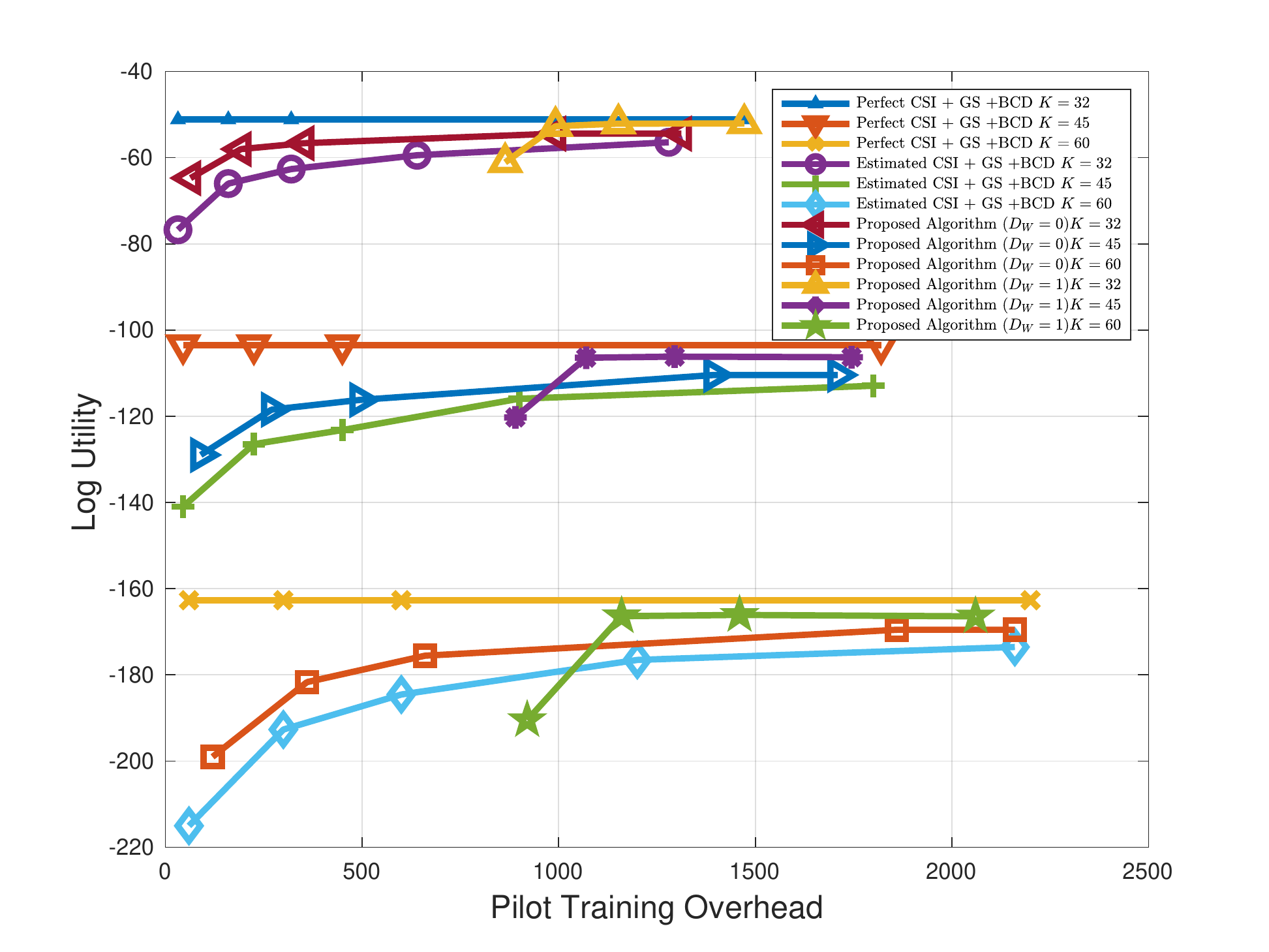}%
\caption{Network-wide log utility of slow-fading channel with different number of users, $N=128$, $M=16$, $P_d=5$dBm.}%
\label{fig.irs_general}%
\end{figure}
\subsection{Generalization to Different Sizes of User Pool}
In an RIS-assisted network, parameters such as the number of antennas at the BS or size of the RIS may be static but the number of users is constantly changing depending on the coverage of the BS and the user traffic.  
Thus, it is important for the proposed algorithm to demonstrate generalizability to scenarios with a different number of users. 

Generalizability in the first two stages is achieved by adopting the same aggregation and combination operation for all user nodes throughout the GNN architectures, i.e., $g_w(\cdot)$, $f_1(\cdot)$, $f_2(\cdot)$, $f_3(\cdot)$, $f_4(\cdot)$, $f_5(\cdot)$ and $\ell_{2M}(\cdot)$, such that the updating rule of the user nodes is independent of the number of users.
Consequently, when the total population of the users changes from $K$ to $K^{\prime}$, the learned aggregation and combination operation can still be applied to the new graph with $K^{\prime}$ user nodes. In the final BS beamforming stage, generalizability is ensured as the BS beamforming optimization is not a function of the total number of users, but a function of the number of BS antennas. In this way, the overall three-stage framework can generalize to systems with a different number of users. 
%\ref{vk_update}

We test the generalizability of the proposed three-stage framework as follows.
We train the GNN on a user pool of size $32$, and test the scheduler's performance in networks where the number of users is $45$ and $60$. 
As seen in Fig.\ref{fig.irs_general}, the proposed algorithms can generalize well. It can converge to near the perfect CSI baseline, and can significantly reduce the pilot overhead as compared to the conventional channel estimation based approach. 

%in a slow-fading setting. In a fast-fading environment, the performance gain also persists when generalizing to systems with different number of users.

\subsection{Discrete vs. Continuous Phase Shifter}

Finally, we test the proposed algorithm on RIS with discrete phase shifts.
In practical implementation, manufacturing RIS elements with infinite precision 
phase shifts may be costly or even infeasible due to hardware limitations\cite{discreteXinT}. %discreteWuQQ
Thus, the optimization of the RIS system with discrete phase shifters is of interest.
To this end, we quantize the continuous reflecting coefficients
by a finite number of levels. 
For an RIS with $b$ control bits, we uniformly divide the range $[0,
2\pi)$ to $2^b$ partitions. The mid-value of each partition is the
representation value of each partition. For example,
%Subsequently, each continuous value is categorized to a unique partition and mapped to the corresponding representation value. 
%$1$ controller bit corresponds to $2^1$ representation values: $\pi/2$ and $3\pi/2$; 
$2$ bits correspond to $2^2$ representation values: $\pi/4$, $3\pi/4$, $5\pi/4$ and $7\pi/4$.

\begin{figure}
\includegraphics[width=\columnwidth]{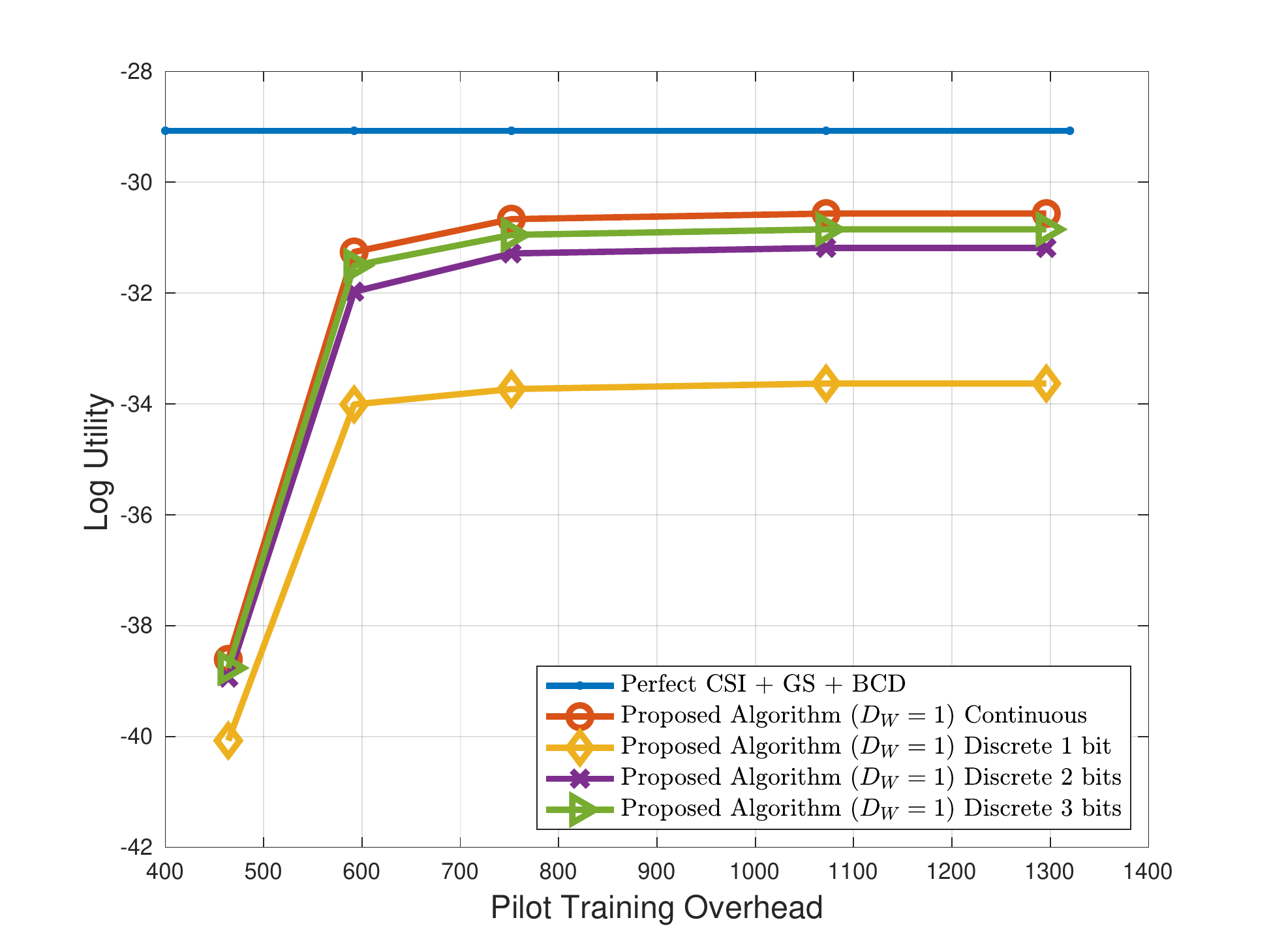}%
\caption{Network-wide log utility of slow-fading channel with different number of control bits, $N=128$, $M=8$, $K=32$, $P_d=15$dBm.}%
\label{fig.contivsdiscrete}%
\end{figure}

Fig.~\ref{fig.contivsdiscrete} shows the performance of the proposed algorithm in log utility with discrete phases at the RIS. It can be observed that 2-bit quantization can already achieve $98\%$ performance of the continuous phase shift case. This implies that with only 2 control bits, an RIS can already deliver a network utility performance similar to an RIS with continuous phase shifts. 

\hfill%

\section{conclusions}
\label{sec.conclusions}

This paper considers a proportional fairness scheduling problem in a
multiuser RIS-assisted MISO network. Specifically, 
%the paper serves as one of the pioneering work in the field of joint scheduling and beamforming in RIS-assisted network under practical setting, where our proposed algorithm
we show that a GNN can simultaneously tackle the challenges in CSI acquisition,
and in designing the proportionally fair scheduling, the optimal RIS phase shifts,
and the beamformers, under limited pilot overhead.  Numerical results show
significant gain over the conventional approach of first explicitly estimating
the channel, then performing network optimization. The proposed approach can be
generalized to scenarios with an arbitrary number of users. It shows considerable
promise of using a machine learning approach for \emph{discrete optimization},
but also points to the importance of judiciously designing the overall framework,
the pilot placement structure, and the appropriate neural network architecture.

\bibliography{references}
\bibliographystyle{IEEEtran}

%\newpage

\end{document}